 \documentclass[final,3p,times]{elsarticle}

\usepackage{amssymb}
\usepackage[table,xcdraw]{xcolor}
\usepackage{tcolorbox}
\usepackage{amsmath,amsfonts}
\usepackage{url}
\usepackage{bm}
\usepackage{multirow}
\usepackage{graphicx}
\usepackage{adjustbox}
\usepackage{tabularx}
\usepackage{longtable}

\renewcommand{\vec}[1]{\boldsymbol{\mathrm{#1}}}

\newcommand{\mydef}{\ensuremath{\triangleq}}
\newcommand{\transp}{\ensuremath{^\mathsf{T}}}

\journal{Computer Speech And Language}

\begin{document}

\begin{frontmatter}




\title{Towards Explainable Spoofed Speech Attribution and Detection:\\a Probabilistic Approach for Characterizing Speech Synthesizer Components}

\author[inst1]{Jagabandhu Mishra}
\author[inst1]{Manasi Chhibber}
\author[inst2]{Hye-jin Shim}

\affiliation[inst1]{organization={School of Computing},
            addressline={University of Eastern Finland}, 
            city={Joensuu},
            postcode={FI-80101}, 
            country={Finland}}

\author[inst1]{Tomi H. Kinnunen}

\affiliation[inst2]{organization={Language Technologies Institute},
            addressline={Carnegie Mellon University}, 
            city={Pittsburgh},
            postcode={15213}, 
            state={PA},
            country={USA}}


\begin{abstract}
We propose an explainable probabilistic framework for characterizing spoofed speech by decomposing it into probabilistic attribute embeddings. Unlike raw high-dimensional countermeasure embeddings, which lack interpretability, the proposed probabilistic attribute embeddings 
aim to detect specific speech synthesizer components, represented through high-level attributes and their corresponding values. 
We use these probabilistic embeddings 
with four classifier back-ends to address two downstream tasks: spoofing detection and spoofing attack attribution. 
The former is the well-known bonafide-spoof detection task, whereas the latter seeks to identify the source method (generator) of a spoofed utterance. We additionally use Shapley values, a widely used technique in machine learning, to quantify the relative contribution of each attribute 
value to the decision-making process in each task. 
Results on the ASVspoof2019 dataset demonstrate the substantial role of 
waveform generator, conversion model outputs, and
inputs in spoofing detection; and inputs, speaker, and duration modeling in spoofing attack attribution. 
In the detection task, the probabilistic attribute embeddings achieve $99.7\%$ balanced accuracy and $0.22\%$ equal error rate (EER), closely matching the performance of raw embeddings ($99.9\%$ balanced accuracy and $0.22\%$ EER). Similarly, in the 
attribution task, our embeddings achieve $90.23\%$ balanced accuracy and $2.07\%$ EER, compared to $90.16\%$ and $2.11\%$ with raw embeddings. These results demonstrate that the proposed framework is both inherently explainable by design and capable of achieving performance comparable to raw CM embeddings.
\end{abstract}


\end{frontmatter}

\section{Introduction}\label{sec:introduction}

Advances in deep generative modeling \cite{Goodfellow2014-GAN} have paved way to flexible generation of high-quality \emph{synthetic media}. Despite the many useful applications, this `\emph{deepfake}' \cite{Verdoliva2020-media-forensics-deepfakes} technology has raised concerns in its exploitation, whether for misinformation spreading in social media, defaming a targeted person, spoofing biometric systems \cite{Ratha2001-enhancing} or conducting advanced social engineering attacks for financial gains, to name a few. Synthetic media targeted specifically against biometric recognition are forms of \emph{presentation attacks} (\emph{spoofing attacks}) \cite{ISO/IEC30107-1:2023}. In this study, our focus is on speech. The two well-known technologies of \emph{speech synthesis} (\text{text-to-speech}, TTS) \cite{tan2021-neural-TTS-survey} and \emph{voice conversion} (VC) \cite{Sisman2021-VC-Overview} can be used for controlled generation and speaker identity conversion of speech utterances, respectively. While the past-generation TTS and VC technology was tailored for small and closed-world speaker populations, recent \emph{zero-shot} \cite{wang2023neural} approaches make it possible, in principle, to reproduce speech in anyone's voice based on a small amount of training data (required at the generation time only). Lately,  a number of alarming social engineering attacks that involve the use of these technologies have been reported \cite{WSJ2019-phone-scam,Arizona2023-kidnapping,cnn2024-hongkong-deepfake}.

Having recognized the potential risks of TTS and VC to spoof voice biometric (speaker recognition) systems from early on \cite{Pellom1999,Masuko1999-HMM-security}, the research community has devised various active defenses (countermeasures) against spoofing attacks. By far, the most common strategy is to \emph{detect} the presence of spoofing:
    \begin{tcolorbox}[colback=blue!5!white,colframe=blue!75!black,title=Spoofing Detection (aka Deepfake Detection or Presentation Attack Detection)]
    \emph{Given a speech utterance \(X\), determine whether \(X\) is more likely to originate from a bonafide human vocal production system (null hypothesis \(H_0\)) or from a spoofing attack algorithm (alternative hypothesis \(H_1\))}.
    \end{tcolorbox}
Apart from a few exceptions \cite{Castan22_odyssey,liu23_speaker_aware_antispoofing}, typically no prior knowledge of the speaker identity in $X$ is assumed. Moreover, in the ASVspoof challenges \cite{Wu2017-ASVspoof}---a widely adapted benchmark in the field with its fifth edition concluded recently \cite{wang2024asvspoof5crowdsourcedspeech}---both the speakers and the spoofing attacks across the training and test data are, to a large extent, disjoint. Such database design choices are intended to promote the development of speaker- and synthesizer-independent spoofing detection suitable for a range of different applications. As for the detection solutions that have emerged through the past decade, the research focus has shifted from hand-crafting features \cite{sahidullah15_feature_comparison,Todisco2017-CQCC} to design of deep neural network architectures \cite{lavrentyeva19_interspeech,Jee2022-AASIST}. To address training data limitations, recent models are typically trained with either data augmentations \cite{Cohen2022-augmentation-antispoofing,Hemlata2022-rawboost} and/or leverage from pre-trained self-supervised feature representations \cite{wang22_self-supervised}. 

The above bonafide-spoofed detectors typically produce a single real-valued detection score (such as estimated likelihood ratio $P(X|H_0)/P(X|H_1)$ or  posterior probability $P(H_0|X)$) that reflects the relative strength of evidence towards the two 
hypotheses. While such scalar score evidence may suffice in applications that involve \emph{automated decision making at scale}---for instance, authenticating millions of users in call-center or automated phone banking---the approach has limited value in applications that require additional level of detail or \emph{explanations} relevant to the user of the technology. Audio forensics is one such domain.

To this end, in this paper, we consider \emph{spoofing attack attribution}, also known as \emph{source tracing}. With the aim to help pinpoint \emph{who} has generated a specific deepfake, it seeks to identify the specific model or algorithm (source) that was used in generating the spoofing attack: 
    \begin{tcolorbox}[colback=blue!5!white,colframe=blue!75!black,title=Spoofing Attack Attribution (aka Source Tracing)]
    \emph{Given a spoofed speech utterance $X$, along with a collection of hypothesized attacks $\mathcal{A}=\{\mathcal{A}_1,\dots,\mathcal{A}_C\}$, determine \emph{which attack} is the source of $X$; or conclude 'none of them' (unknown attack not included in $\mathcal{A}$).}
    \end{tcolorbox}
When we know that $X$ actually originates from one of the attacks in $\mathcal{A}$, the spoofing attack attribution is said to have a \emph{closed set} of spoofing attacks; otherwise an \emph{open set}. 
Our study is not the first to address attack attribution, either in general multimedia~\cite{liu2025model,wang2023evaluating} 
or in speech (related studies are reviewed in Section \ref{Sec:related-work}). Existing approaches in the literature are primarily 
focused on predictive performance rather than interpretability. In contrast, we propose a novel \textbf{probabilistic attribution} framework (detailed in Section \ref{sec:proposed-probabilistic-characterization}) which is designed to be inherently explainable. An initial study 
this method has been reported in our recent work~\cite{chhibber2024explainable}, where the study was limited to using a classic decision tree classifier~\cite{breiman2017classification}. While decision trees provided a degree of interpretability, their predictive performance can be 
constrained. 
Moreover, the earlier work addressed the spoofing attack attribution task using only two 
attacks. 
The present study substantially expands upon ~\cite{chhibber2024explainable} both at the conceptual, theoretical framework, and experimental levels. 
We incorporate a diverse range of interpretable back-end classifiers, including naive Bayes~\cite{bishop2006pattern} (specifically tailored to handle probabilistic attribute features), logistic regression~\cite{mccullagh2019generalized}, and support vector machines (SVM)~\cite{cortes1995support}. These classifiers bring complementary strengths, such as probabilistic reasoning (naive Bayes), robust classification (logistic regression), and the ability to handle decision boundaries with maximum margin for improved generalizability (SVM). This diversification enhances both the interpretability and predictive performance of the framework. Furthermore, to achieve better generalizability, this study introduces a new custom protocol on the ASVspoof2019 dataset, expanding the scope beyond the earlier focus on two known attack types to include all $17$ types of attacks for spoofing attack attribution.



\begin{figure}[t!]
    \centering
    \includegraphics[height= 240pt,width=390pt]{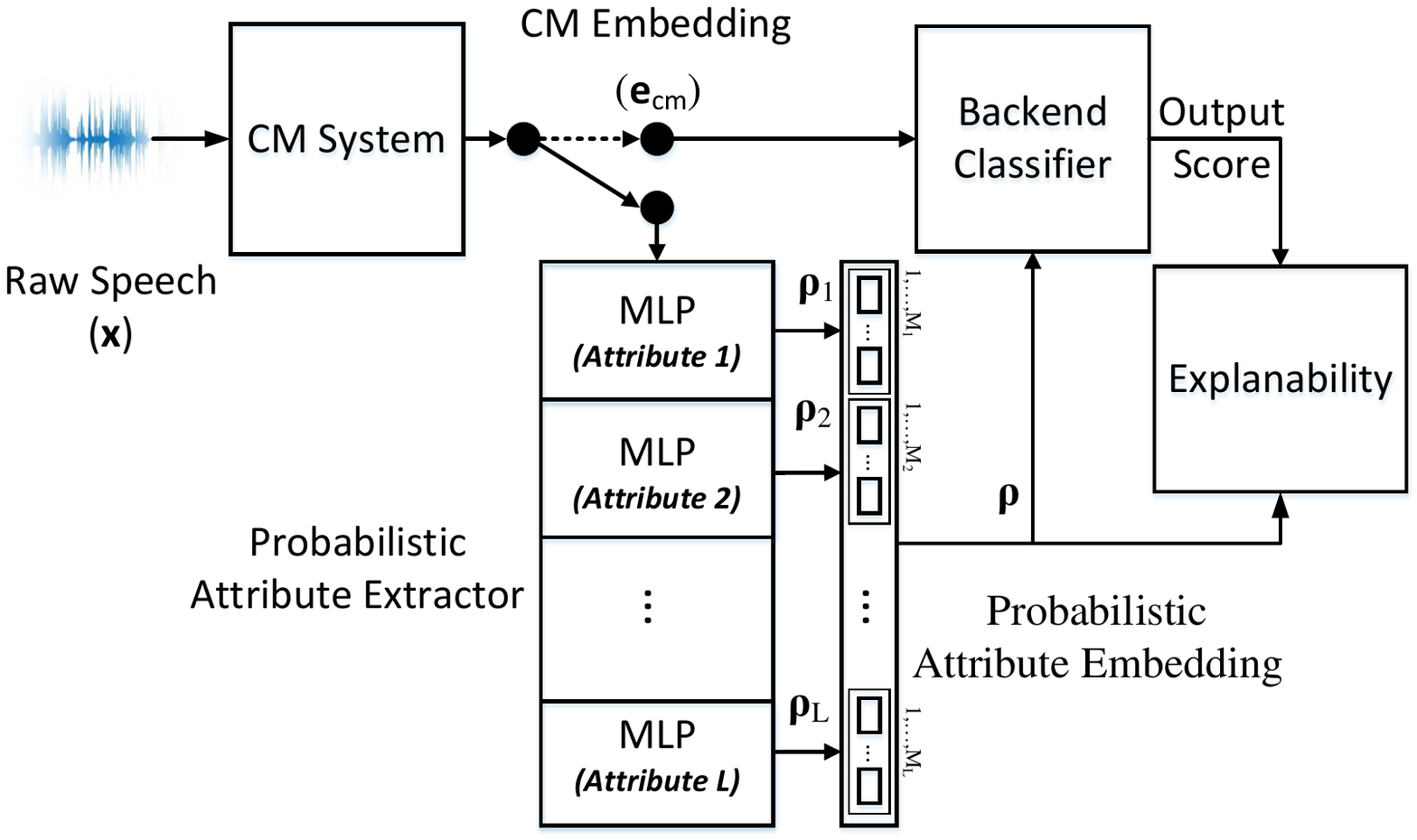}
    \vspace{-1.0 cm}
    \caption{Overall pipeline of the proposed architecture. CM: countermeasure, MLP: multilayer perception. Here $1,\ldots, M_l$ represents the predicted attribute values of attribute $l$. The CM embeddings are extracted from the speech signal and subsequently used with probabilistic attribute extractors to obtain the probabilistic attribute embeddings. The probabilistic attribute embeddings are then used with backend classifiers to perform the downstream tasks: spoofing detection and spoofing attack attribution. For comparison, the CM embeddings are also directly used with backend classifiers to perform the downstream tasks (as shown in Figure through dotted arrow). Finally, the explainable probabilistic attribute embeddings, interpretable classifier, and their scores are used to explain the relative contribution of the attributes in decision-making.}
    \label{Blk_diag_overall}
\end{figure}
\begin{figure}[!t]
    \centering
     \fbox{\includegraphics[height= 215pt,width=460pt]{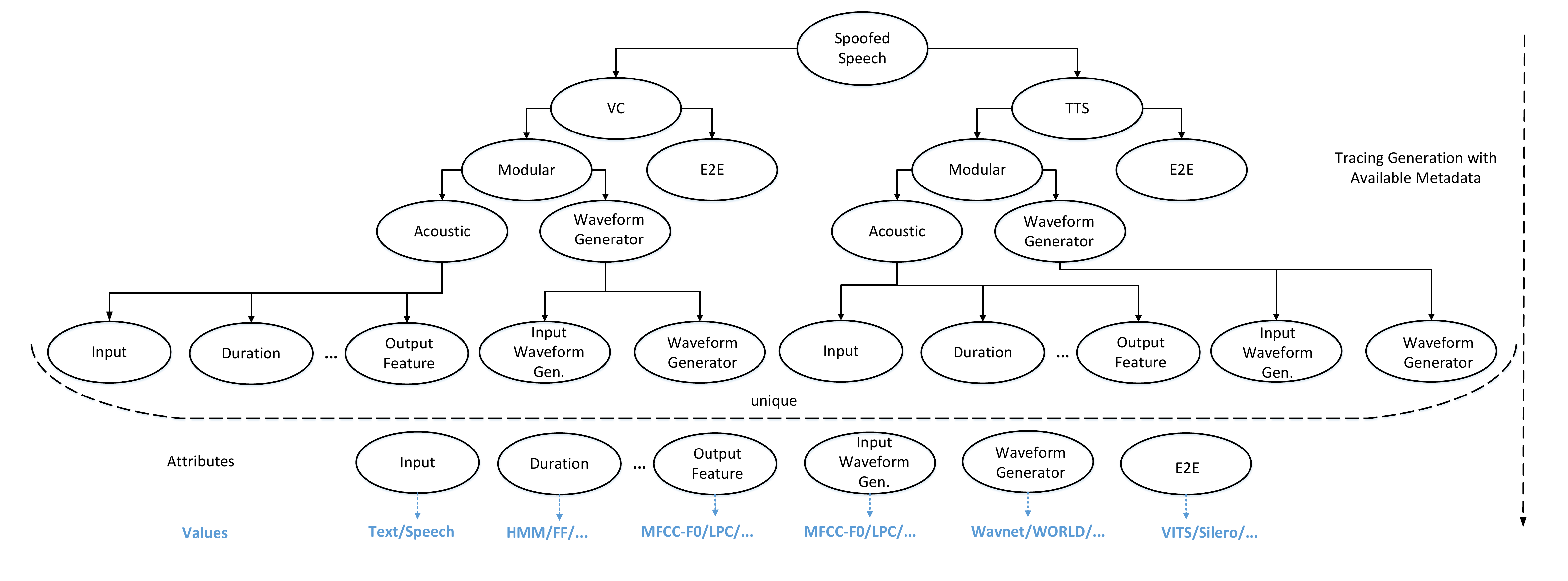}}
    \caption{A tentative structure of spoofed speech generation, VC: voice conversion, E2E: End to End, TTS: text to speech, Waveform Gen.: Waveform Generator}.
    \label{fig:spf_gen_tree}
\vspace{-0.5cm}
\end{figure}

The central idea (illustrated in Figure~\ref{Blk_diag_overall}) is to extract discrete probability distributions over attribute values during the test (inference) phase, which serve as recording-level feature vectors to address the downstream tasks of spoofing detection and spoofing attack attribution. These attribute values are derived from a compact set of attributes that characterize various spoofed speech generation methods (see Figure~\ref{fig:spf_gen_tree}). 
For a given test utterance, the explainable design provides insights in the form of probabilities associated with the attributes involved in the spoofed speech generation process. Additionally, while \emph{Shapley values}~\cite{scott2017unified} in existing literature have been used to analyze contributing factors in the speech signal or spectrogram for spoofing detection~\cite{ge2022explaining}, our work extends this approach to analyze the contribution of specific attributes towards decision-making. This extension enhances the interpretability of the proposed framework by identifying the key attributes influencing the outcomes for both spoofing detection and spoofing attack attribution tasks. 

\section{Related work: inspirations from binary attribution, and studies on explainable deepfake characterization}
\label{Sec:related-work}


In this study, our intention is to demonstrate how different spoofing attack algorithms can be broken down into their elementary building blocks, represented through a high-level probabilistic feature description. 
 Specifically, our aim is to characterize 
spoofing attacks through abstract \textbf{discrete probabilistic attributes}. As for the \emph{`discrete'}, random events that can be named, enumerated, and explained to others are convenient to us as humans. Much of the basic sciences (and everyday communication) seeks to explain complicated concepts through simplifications. 
As for the \emph{`probabilistic'}, real-world physical objects, including acoustic speech waveforms, are rarely `either-this-or-that' due to inaccuracies in measuring or observing the relevant cues; probabilities provide a natural way to quantify uncertainty in the features, allowing one to postpone  the final classification decision to a later stage. 
Finally, \emph{`attribute'} is essentially a synonym for `feature', but intended to emphasize the notion of the features to bear clear semantics to humans.

The prior work closest in sentiment to our work includes the use of \emph{binary attributes} \cite{Bonastre2011-binary-speaker,benamor24_interspeech,Imen2024-PhD-thesis} for speaker characterization. By motivations stemming partly from presentation of DNA evidence in forensic contexts \cite{amor2024deep}, this approach uses binary-valued (`on/off') attributes extracted either through classic statistical GMM \cite{Bonastre2011-binary-speaker} or modern neural autoencoders \cite{benamor24_interspeech} to characterize speaker-specific features. We point the interested reader to~\cite{Imen2024-PhD-thesis} for in-depth motivations and interpretation 
of this binary attribution approach, designed to improve explainability in forensic speaker profiling. These studies provide a valuable foundation for feature-based analysis in speech, highlighting the potential of interpretable representations, especially within domains like forensics that demand explanations along with accuracy. However, the challenge of spoofing detection demands new strategies that can go beyond individual speaker traits. Whereas our study shares the 
sentiment of \cite{Bonastre2011-binary-speaker,benamor24_interspeech,Imen2024-PhD-thesis} in the desire to define explainable attributes, our study is different 
both in the assumed format of the attributes and the classification tasks. Unlike \cite{Bonastre2011-binary-speaker,benamor24_interspeech,Imen2024-PhD-thesis} which rely on binary-valued (`on/off') attributes, our method uses multi-class valued attributes and represents feature value uncertainty with probabilities rather than binarized ($0$ or $1$) versions. This allows such probabilistic features to be easily combined with arbitrary classifiers (as we demonstrate in our experiments). This move to a probabilistic multi-class representation allows capturing rich nuances in spoofing signals, which might otherwise be lost in hard binary decisions.

Apart from the subtle technical modeling choice on attribute format, the more critical difference is underlined in the task (speaker characterization in  \cite{Bonastre2011-binary-speaker,benamor24_interspeech,Imen2024-PhD-thesis}; deepfake characterization in the present work). Speaker characterization leverages audio data that originates from a biological system---the human vocal production system driven by neural and muscular activities that consists of the lungs, the larynx, and the supralaryngeal vocal tract system. As such, motivation and design of attributes for speaker characterization can benefit from the established speech science 
that covers domains such as physiology, biomechanics, aerodynamics, acoustics, digital speech modeling, phonetics and phonology. In contrast, deepfakes originate from artificial systems (computer programs that implement specific synthesis or conversion models). What are the relevant attributes of such systems --- how should we characterize spoofed and deepfake speech? Whereas the field of speech synthesis 
has a long history and its 
established system categorizations (e.g. concatenative, statistical, and neural TTS), there has been 
little focus towards the attribution task. In short, we require thorough re-thinking of what the explainable attributes for categorizing, classifying and explaining different spoofing attack algorithms should be. With deepfakes being still a relatively recent security threat, and with major focus of defense solutions having been on the binary 'real-fake' discrimination task, the work in 
more refined categorization of spoofed speech generators (including their detection) is currently taking only its infant steps. By no means the present work is intended to be a full description of spoofed speech characterization but, rather, a modest conceptual step forward. Besides forensic `speaker profiling', an already established topic \cite{campbell2009forensic}, we expect future forensic speech analysts to benefit from `deepfake profiling' or `attack fingerprinting' methodologies. To the best of the authors' knowledge, however, no such standardized approaches are available yet.

\begin{table}[h!]
\centering
\caption{Review of speech-based deepfake attribution and detection studies. This table highlights the variety of approaches employed, including the feature extraction and classification methods used by different researchers.}
\label{tab:review}
\refstepcounter{table}
\label{tab:models_accuracy}
\resizebox{\columnwidth}{!}{
\begin{tabular}{|l|l|l|l|l|l|}  
\hline
\textbf{Paper} & \textbf{Task}                      & \textbf{Dataset}             & \textbf{Features} & \textbf{Classifiers}     & \textbf{Methodology}                              \\ 
\hline
          \cite{albadawy2019detecting}     & Attribution                 & In-house data    & \begin{tabular}[c]{@{}l@{}}bicoherence magnitude and phase,\\first four statistical moments of\\magnitude and phase\end{tabular} & \begin{tabular}[c]{@{}l@{}}logistic regression (linear classifiers)\end{tabular}  & \begin{tabular}[c]{@{}l@{}}Bispectral analysis was used to extract features\\highlighting unusual spectral correlations in\\spoofed speech for classification against human\\speech.\end{tabular}                               \\ 
\hline
         \cite{borrelli2021synthetic}      & Attribution                 & ASVspoof2019        & \begin{tabular}[c]{@{}l@{}}short-term and long-term prediction\\traces (obtained by minimizing prediction\\error in the time domain)\end{tabular} & \begin{tabular}[c]{@{}l@{}}linear SVM, RBF SVM, random\\forest, GMM\end{tabular} & \begin{tabular}[c]{@{}l@{}}A method is proposed that combines short-term\\and long-term prediction error energies, extracted\\from a source-filter model of speech, to classify\\audio as either real or synthetic speech.\end{tabular}             \\ 
\hline
           \cite{salvi2022exploring}    & Attribution                 & ASVspoof2019         & \begin{tabular}[c]{@{}l@{}}spectrograms, MFCCs, CQCCs, LFCCs,\\raw waveforms\end{tabular} & ResNet, RawNet2, GMM & \begin{tabular}[c]{@{}l@{}}Several different audio features were extracted\\and used with various classifiers (SVM, GMM, and\\ResNet) to detect synthetic speech in both closed\\and open-set scenarios, comparing their performance.\end{tabular}                 \\ 
\hline
            \cite{rahman2023syn}   & Attribution                 & \begin{tabular}[c]{@{}l@{}}SP Cup challenge\\(DARPA-Semafor)\end{tabular} & log-mel spectrograms & \begin{tabular}[c]{@{}l@{}}multi-layer CNNs (used to extract\\features and classify), self-attention\\layers (used by transformer in feature\\extraction)\end{tabular} & \begin{tabular}[c]{@{}l@{}}A semi-supervised multi-class ensemble of CNNs\\was trained using a novel unknown class strategy,\\data augmentation, and ensemble techniques for\\synthetic speech attribution.\end{tabular}                     \\ 
\hline
            \cite{bhagtani2023synthesized}   & Attribution                 & \begin{tabular}[c]{@{}l@{}}ASVspoof2019,\\DARPA-Semafor\end{tabular}   & mel-spectrograms &   PSAT & \begin{tabular}[c]{@{}l@{}}A patchout spectrogram attribution transformer\\(PSAT) architecture, combining a patchout\\mechanism with a transformer network, was\\trained to classify synthesized speech samples\\and attribute them to their respective synthesis\\algorithms.\end{tabular}                  \\ 
\hline
           \cite{bartusiak2022transformer}    & Attribution                 & DARPA-Semafor            & spectrograms & compact attribution transformer (CAT) & \begin{tabular}[c]{@{}l@{}}A compact attribution transformer (CAT) was\\trained to classify speech signals into known\\synthesizers or an unknown class, utilizing\\normalized spectrograms as input.\end{tabular}  \\ 
\hline
        \cite{yadav2023synthetic}       & Attribution                 & \begin{tabular}[c]{@{}l@{}}ASVspoof2019,\\DARPA-Semafor\end{tabular}   & \begin{tabular}[c]{@{}l@{}}mel-spectrogram, 768-dim. representation of\\each region of spectrogram (using pretrained\\transformer), positional encodings\end{tabular} &  \begin{tabular}[c]{@{}l@{}}linear layer with SoftMax activation\\(on top of mean pooled representation\\from pretrained transformer)\end{tabular}  & \begin{tabular}[c]{@{}l@{}}A  synthetic speech
attribution transformer\\(SSAT) was trained using a pretext task to learn\\robust representations of speech signals, which\\were then used for synthesizer attribution in both\\closed and open set scenarios.\end{tabular}           \\ 
\hline
          \cite{muller2022attacker}     & Attribution                 & ASVspoof2019     & \begin{tabular}[c]{@{}l@{}}low-level acoustic features, neural\\embeddings\end{tabular} & \begin{tabular}[c]{@{}l@{}}feed-forward neural network (used\\for both low-level and learned\\embedding based signatures)\end{tabular} & \begin{tabular}[c]{@{}l@{}}Low-level audio features are extracted and used\\to train a classifier for attributing audio deepfakes\\to specific attacks.\end{tabular}          \\ 
\hline
           \cite{klein24_interspeech}    & Attribution                 & \begin{tabular}[c]{@{}l@{}}ASVspoof2019,\\MLAAD\end{tabular}  & \begin{tabular}[c]{@{}l@{}}raw audio, LFCC (with delta and double\\delta)\end{tabular} &       \begin{tabular}[c]{@{}l@{}}ResNet, self-supervised learning\\(SSL) based model using wav2vec\\2.0, Whisper based model\end{tabular}      & \begin{tabular}[c]{@{}l@{}}A two-stage system is used, first training\\a model for speaker classification using\\self-supervised learning and then fine-tuning\\it for source tracing using a multi-task learning\\approach.\end{tabular}               \\ 
\hline
            \cite{zhu2022source}   & \begin{tabular}[c]{@{}l@{}}Detection and \\Attribution\end{tabular} & ASVspoof2019    & \begin{tabular}[c]{@{}l@{}}log-FBank, truncated/concatenated signals\\(for RawNet2), various acoustic features\\(e.g. LPCC, MFCC - used as inputs, no \\ explicit feature extraction layer)\end{tabular} &  \begin{tabular}[c]{@{}l@{}}ResNet34, RawNet2 (used as front-\\end with classification layers)\end{tabular}     & \begin{tabular}[c]{@{}l@{}}A multi-task learning strategy is employed,\\combining speaker classification and spoofing\\detection to improve source tracing performance.\end{tabular}                               \\ 
\hline
         \cite{zeng2023deepfake}      & Attribution                 & \begin{tabular}[c]{@{}l@{}}ADD2023 Challenge\\Track 3 dataset\end{tabular} & wav2vec2.0 & ECAPA-TDNN & \begin{tabular}[c]{@{}l@{}}A system using wav2vec2.0 for feature extraction,\\ECAPA-TDNN for algorithm classification, and\\data augmentation techniques.\end{tabular}                                \\ 
\hline
           \cite{tian2023deepfake}    & Attribution                 & \begin{tabular}[c]{@{}l@{}}ADD2023 Challenge\\Track 3 dataset\end{tabular} & LFCC, Raw Waveform, HuBERT & rawnet2, SE-Res2Net50, HuBERT & \begin{tabular}[c]{@{}l@{}}Multi-model fusion using manifold distance and\\feature-level fusion.\end{tabular}                                \\ 
\hline
      \cite{lu2023detecting}         & Attribution                 & \begin{tabular}[c]{@{}l@{}}ADD2023 Challenge\\Track 3 dataset\end{tabular} & STFT, Wav2Vec2  & \begin{tabular}[c]{@{}l@{}}SENet, LCNN-LSTM, TDNN,\\kNN-based OOD Detection\end{tabular} & \begin{tabular}[c]{@{}l@{}}kNN with cosine similarity for OOD detection, \\combined with data augmentation and model fusion.\end{tabular}                               \\
\hline
      \cite{liu24m_interspeech}         & Detection                 & \begin{tabular}[c]{@{}l@{}}PartialSpoof dataset,\\ASVspoof 2019 LA dataset\end{tabular} & log-mel spectrograms  & SSL-gMLPs, SSL-Res1D & \begin{tabular}[c]{@{}l@{}}Grad-CAM visualization and a novel quantitative\\analysis metric (RCQ) were used to interpret the\\decision-making process of countermeasures\\trained on partially spoofed audio.\end{tabular}                               \\
\hline
      \cite{gupta24b_interspeech}         & Detection & \begin{tabular}[c]{@{}l@{}}LJSpeech dataset, AI-generated\\voice data using Tacotron2 and\\FastSpeech2 with HiFiGAN\\ waveform generator\end{tabular} & mel-spectrogram & CNN14 & \begin{tabular}[c]{@{}l@{}}PDSM discretizes saliency maps using phoneme\\boundaries from an ASR model for explainable\\spoofed speech detection.\end{tabular}                              \\
      
\hline
\cite{zhu2022source} &
  \begin{tabular}[c]{@{}l@{}}Detection and \\ Attribution\end{tabular} &
  ASVspoof2019 &
  raw audio &
  Resnet34, Rawnet2 &
  \begin{tabular}[c]{@{}l@{}}A multitask framework is introduced to jointly train \\ spoofing detection and attribute value classification.\end{tabular} \\ \hline
\cite{phukan2024investigating} &
  Attribution &
  \begin{tabular}[c]{@{}l@{}}ASVspoof2019 and FAD\\ 2023\end{tabular} &
  features from pre-trained models &
  \begin{tabular}[c]{@{}l@{}}CNN, FusIon through ReNyi\\  DivERgence (FINDER)\end{tabular} &
  \begin{tabular}[c]{@{}l@{}} FINDER framework is proposed for optimally\\ combine the features from speech-pre-trained models\end{tabular} \\ \hline
\cite{zhang2025transformer} &
  Attribution &
  \begin{tabular}[c]{@{}l@{}}ASVspoof2019, DARPA SemaFor,\\  and CFAD\end{tabular} &
  Sub-band time-frequency feature &
  Transformer &
  \begin{tabular}[c]{@{}l@{}}Proposed sub-band time-frequency feature and shown\\ the effectiveness of class-dependent threshold for \\ openset attribution task.\end{tabular} \\ \hline
\cite{xie2025neural} &
  Attribution &
  ST-Codecfake &
  raw audio, mel spectrogram &
  Wav2vec ASSIST, ASSIST, LCNN &
  \begin{tabular}[c]{@{}l@{}}Introduced a source tracing Codecfake dataset, and \\ analyzed the performance with frequently used\\ classifiers.\end{tabular} \\ \hline
\textbf{This work} &
  \begin{tabular}[c]{@{}l@{}}Detection and \\ Attribution\end{tabular} &
  \textbf{ASVspoof2019} &
  \begin{tabular}[c]{@{}l@{}}probabilistic attribute embeddings\\ derived from countermeasure systems\end{tabular} &
  \begin{tabular}[c]{@{}l@{}}Naive Bayes, Decision Tree, Logistic\\ Regression, Support Vector Machine\end{tabular} &
  \begin{tabular}[c]{@{}l@{}}Spoofed speech is characterized by a set of high-\\ level probabilistic attributes extracted from CM\\ embeddings and used with several classifiers,\\ including Shapley value analysis for explainability.\end{tabular} \\ \hline
\end{tabular}%
}
\end{table}

With the stage set above for explainability and the relevance of probabilistic attributes in speech tasks, we now review related explainability studies specifically within the context of spoofing detection and spoofing attack attribution, as summarized in Table~\ref{tab:review}. These studies are examined through the lens of their focus on feature analysis, detection or attribution, and modeling techniques. Prior works on explainability in this domain can be broadly categorized into two approaches. The first category focuses on identifying regions within the spectrogram or speech signal that contribute to predictions in spoofing detection. For instance,~\cite{liu24m_interspeech} employs gradient-weighted class activation mapping (Grad-CAM), while~\cite{gupta24b_interspeech} uses saliency maps to highlight the regions of speech waveforms and spectrograms contributing to predictions. Although these approaches provide a degree of explainability, they are not explicitly linked to the underlying process of spoofed speech generation.

The second category attempts to explain spoofed speech by tracing its source of generation, commonly referred to as \emph{spoofing attack attribution} or \emph{source tracing}, as mentioned in Section \ref{sec:introduction}. One of the earliest investigations in this direction \cite{albadawy2019detecting} utilized bispectral magnitude and phase features to differentiate between spoofed and bonafide speech sources. Building on this, \cite{borrelli2021synthetic} explored long-term and short-term parameterization using linear prediction (LP) residuals and their periodicity to detect spoofing attributes, highlighting the importance of time-domain feature analysis. Open-set attribution was investigated in \cite{salvi2022exploring} through a one-class thresholding approach combined with a support vector machine (SVM). Similarly, \cite{rahman2023syn} addressed open-set attribution using an ensemble semi-supervised multi-class strategy based on convolutional neural networks (CNNs). Further advances in spoofing attack attribution include approaches based on the popular Transformer models. For instance, \cite{bhagtani2023synthesized} proposed a patchout spectrogram-based Transformer framework, while \cite{bartusiak2022transformer} introduced a compact attribution transformer with a poly-1 loss function \cite{leng2022polyloss} to enhance open-set performance. Extending this direction, \cite{yadav2023synthetic} explored self-supervised Transformer models for spoofing attack attribution. Another study \cite{muller2022attacker} examined low-level acoustic features such as fundamental frequency, distortion, shimmer, and speaker gender, combined with neural embeddings to improve attack attribution. Moreover, \cite{phukan2024investigating} demonstrated the role of prosodic signatures captured in speech pre-trained models, while \cite{zhang2025transformer} explored sub-band time-frequency features with Transformers for spoofing attack attribution. Recently, \cite{klein24_interspeech} employed Rawnet \cite{jung2019rawnet}, Whisper \cite{huang1995microsoft}, and self-supervised learning techniques \cite{klein2024source} to trace spoofed speech sources, focusing on attributes such as input type, acoustic model, and waveform generator \cite{klein2024source}. Another approach in \cite{zhu2022source} utilized Rawnet and ResNet34 architectures with multi-task learning (joint attribute value classification and spoofing detection) to address source tracing alongside spoofing detection. Furthermore, \cite{xie2025neural} introduced neural codec source tracing and demonstrated its effectiveness in both open-set and closed-set scenarios, leveraging mel light CNNs (mel-LCNN)~\cite{wu20c_interspeech}, audio anti-spoofing using integrated spectro-temporal graph attention network (AASIST)~\cite{jung2022aasist}, and wav2vec-AASIST~\cite{tak2022automatic} architectures. The growing interest in this field is evident from the upcoming special session on \emph{Source tracing: The origins of synthetic or manipulated speech}\footnote{\url{https://www.interspeech2025.org/special-sessions}} at Interspeech 2025, which aims to bring together researchers working on spoofing attribution techniques. Additionally, as summarized in Table~\ref{tab:review}, most prior works utilized publicly available datasets such as ASVspoof2019 \cite{wang2020asvspoof}, Defense Advanced Research Projects Agency Semantic Forensics (DARPA SemaFor) \cite{darpa}, TIMIT-TTS \cite{salvi2023timit}, Open Dataset of Synthetic Speech (ODSS) \cite{yaroshchuk2023open}, Fake or Real (FoR) \cite{reimao2019dataset}, Chinese fake audio detection (CFAD) \cite{MA2024103122} and Multi-Language Audio Anti-Spoof (MLAAD) \cite{muller2024mlaad} for evaluating the performance of their methods.

Unlike speaker profiling, which focuses on analyzing speaker traits, most existing techniques prioritize predictive performance in either detecting spoofed speech or attributing attacks but offer a limited explanation of the spoofing traits that contribute to the predictions. In contrast, our approach characterizes spoofed speech using discrete probabilistic attributes derived from its generation process. These attributes are leveraged to perform both spoofing detection and spoofing attack attribution.  By design, the proposed framework provides explainability for a given utterance by associating probabilities with the attributes involved in the spoofed speech generation process, offering insights into both tasks simultaneously.



\section{Proposed Probabilistic Attribute Characterization}\label{sec:proposed-probabilistic-characterization}



In our approach, a CM embedding ($\Vec{e}_\text{cm}$) is first extracted from a speech utterance ($\Vec{x}$), as illustrated in Figure~\ref{Blk_diag_overall}. These CM embeddings are then utilized to characterize the attributes through probabilistic attribute extractors. The outputs of each attribute extracted are stacked to form a probabilistic attribute embedding, denoted as $\Vec{\rho}$. These embeddings are subsequently used with backend classifiers to perform downstream tasks here, either of spoofing detection or spoofing attack attribution. To enhance explainability, the attributes that contribute significantly to the predictions are analyzed using Shapley values \cite{rozemberczki2022shapley}. For a fair performance comparison, the downstream tasks are also conducted using raw CM embeddings as baselines.

The design of attributes, their probabilistic characterization, the use of backend classifiers for downstream tasks, and the use of Shapley values to explain attribute contributions to decision-making are briefly discussed in the following subsections.


\subsection{Attributes of spoofed speech generators}
\label{sec:prob_att_char}



We aim to characterize spoofed speech using explainable probabilistic attribute embeddings, drawing inspiration from the processes involved in its generation. To this end, we decompose 
spoofed speech 
generation to its 
underlying attributes hierarchically, as illustrated in Figure \ref{fig:spf_gen_tree}. First, note that spoofed utterances are typically produced through \emph{voice conversion} (VC) or \emph{text-to-speech} (TTS) approaches, or their hybrids (TTS followed by VC)~\cite{zhu2022source,klein24_interspeech,wang2020asvspoof}. These methods are generally designed and implemented through one of the two main approaches. The first approach combines a number of clearly identified sub-components---such as an intermediate acoustic feature predictor and a waveform generator---in a cascade. Since the individual sub-components (modules) and their functions are clearly understood and known at the training and inference (speech generation) stages, we refer to spoofed speech generators in this category \textbf{modular}. This 
is in contrast to the second approach based on  
\textbf{\emph{end-to-end}} (E2E)~\cite{salvi2023timit} models that 
map either a text passage (in TTS) or a source speaker utterance (in VC) directly to a predicted waveform. 
The design philosophy of E2E approaches is to use a model optimized `in one go', as opposed to constructing it from separately-optimized sub-components. Note that while E2E models also consist of cascaded components---namely, the layers in a deep neural network that pass intermediate latent feature representations to the next layer(s)---the key distinction here is whether the individual sub-components have a clearly-identified, explainable meaning to their function or not. A 'WaveNet waveform generator', for instance, might be argued to be more meaningful than `layer number 26'.

The modules used in constructing a complete TTS or VC system themselves can further be characterized with a number of pre-defined high-level design choices, or \emph{attributes}. The \emph{values} of attributes determine the design choices used. For instance, the intermediate acoustic feature predictor operates on specific type of input data (text or speech) and produces particular type of outputs (type of predicted features) expected by the subsequent waveform generator module. The intermediate acoustic feature predictor may use a specific duration model or a particular target speaker representation (such as one-hot vector or an external speaker embedding vector for zero-shot type of TTS). Assigning specific values for both the feature predictor and the waveform generator attributes defines a particular TTS or VC system design. 

As a concrete example, consider one of the widely used modular TTS approaches described in 
~\cite{shen2018tacatron2}. This approach consists of a so-called 
`Tacotron2' acoustic feature predictor and a `WaveNet' waveform generator~\cite{van2016wavenet}. The 
former assumes character embedding inputs that are passed through layers of convolutional neural network, long-short term memory (LSTM), and bidirectional LSTM layers to predict a mel-spectrogram. A WaveNet waveform generator conditioned on the mel-spectrogram 
is then used for generating the audio waveform. This approach can further be extended, as reported in ~\cite{cooper2020zero}, to support multiple speakers by conditioning the mel-spectrogram generation by speaker embeddings.   
 Hence, based on the structure shown in Figure~\ref{fig:spf_gen_tree}, the `Tacotron2+WaveNet' TTS system can be characterized using the following attribute-value pairs:
    \begin{verbatim}
        input                                = text
        input processor(character encoder)   = 1-hot
        duration                             = None
        conversion                           = CNN+LSTM
        speaker                              = x-vector
        outputs                              = mel-spectrogram
        waveform generator                   = waveNet
    \end{verbatim}

This way, we can characterize the spoofed speech generation methods using information available at the training time. Whereas the majority of source tracing (attack attribution) approaches aim at fingerprinting the \textbf{full} model, our approach seeks to detect \textbf{partial} traits by estimating the attribute values. In practice, each attribute is presented using a categorical random variable over its possible values. While these values are known at the training time (presented through 1-hot encoding), their values are uncertain at the test time.
We hypothesize that variations in the attribute values 
may leave detectable traces in the generated spoofed speech. 
Compared to the typical high-dimensional CM embeddings, the characterization of spoofed speech with clearly-defined attributes and their values has better explainability. 

\subsection{Probabilistic attributes}

Let $\vec{x}$ denote a speech waveform. We use ${\bm \rho}_{\ell}( \Vec{x})=[\rho_{\ell,1},\dots,\rho_{\ell,M_\ell}]\transp$ to denote a discrete probability distribution over the values of attribute $\ell$; the probabilities are $\rho_{\ell,m}( \Vec{x})$, where 
$1 \leq m \leq M_\ell $. For a given utterance, we extract $L$ such distributions. 
In our approach, we first extract a CM embedding ${\bf e}_{\text{cm}}(\Vec{x}) \in \mathbb{R}^D$ using a suitable spoofing CM system. The  CM embeddings are then used to train a bank of $L$ \emph{probabilistic attribute extractors}, $\{\mathcal{F}^{\text{pae}}_{{\bm \rho}_{\ell}}: \ell=1,\dots,L\}$, each specific to an attribute as follows,
 \begin{equation}
     \mathcal{F}^{\text{pae}}_{{\bm \rho}_{\ell}}:{\bf e}_{\text{cm}}( \Vec{x}) \mapsto {\bm \rho}_{\ell}( \Vec{x}) \in \mathbf{P}^{M_\ell},
 \end{equation}
where 
$\mathbf{P}^{K} \mydef \{(r_1,\dots,r_K) : r_k \geq 0\; \sum_{k=1}^{K} r_k = 1\}$
denotes a probability simplex. The $L$ stacked (concatenated) 
attribute probability distributions, 
    \begin{equation}
        \vec{\rho}(\vec{x})\mydef [\vec{\rho}_1(\vec{x}), \vec{\rho}_2(\vec{x}), \ldots,\vec{\rho}_L(\vec{x})]\transp \in \mathbf{P}^{M_1}\times\dots \times \mathbf{P}^{M_L}, 
    \end{equation}
of dimensionality $M = \sum_{\ell=1}^L M_\ell$, will be referred to as the \textbf{probabilistic attribute embedding} of the utterance $\vec{x}$. Labeled collections of these embeddings can then be treated like any generic features to train the back-end classifier for relevant downstream tasks. Compared to the original (raw) CM embedding, the dimensionality of the probabilistic attribute embedding is usually lower (i.e. $M < D$). Compared to the abstract raw CM embedding values, in turn, the semantics of the extracted attribute values are clear from the context (determined by the metadata labels used for training the probabilistic attribute extractors). The details of the attribute extractors (implemented through simple multi-layer perceptrons~\cite{yegnanarayana2009artificial}) are provided in the experimental set-up part (Section \ref{sec:exp-setup}). In the experimental results part (Section \ref{sec:exp-results}) we compare the discrimination capability of the original CM embeddings and the probabilistic attribute embeddings.  

\subsection{Back-end classifiers}
\label{section:downstram-task}

In principle, 
any off-the-shelf classifier can be used to address the spoofing detection and attribution tasks. While our experiments includes results using four different approaches, 
we provide here a description of one of them, namely, a \textbf{naive Bayes} classifier. 
Since our features themselves have a probability interpretation, it is useful to provide 
relevant details of this standard probabilistic classifier so as to avoid conflating probabilistic features and probabilistic classifiers. 

As a multi-class task, let us consider the closed-set spoofing attack attribution task as our running example. We are given an unseen speech waveform $\vec{x}$, already known to be a deepfake, but with its source (attack algorithm) remaining unknown. Following the closed-set attack assumption, $\vec{x}$ is hypothesized to have been generated by exactly one of the $C$ known sources (attack algorithms). Since the attack algorithm is not known with certainty, we associate a discrete random variable $\mathcal{A} \in  \{\mathcal{A}_1,\dots,\mathcal{A}_C\}$ to the state of the variable associated with the attack label. Under the assumption that both the \emph{prior probabilities} $\pi_i \mydef P(\mathcal{A}=\mathcal{A}_i)$ and the \emph{likelihood functions} $P_i(\vec{x})\mydef P(\vec{x}|\mathcal{A}=\mathcal{A}_i)$ are known, it is well known that the posterior probability given by the \emph{Bayes' rule},
\begin{equation}
    P(\mathcal{A}=\mathcal{A}_i|{\bf x})=\frac{\pi_i\, P_i({\bf x})}{P({\bf x})},
\end{equation}
forms the basis for optimal decisions (decisions that minimize classification error probability). Here, 

\begin{equation}
 P(\vec{x})\mydef\sum_{j=1}^N \pi_j P_j(\vec{x}),   
\end{equation}
is a normalizer that ensures that a valid posterior probability distribution results. Given the initially uncertain knowledge of the state $\mathcal{A}$, the Bayes' rule provides a way to update the knowledge of the attack label. The prior $\pi_i$ and the likelihood $P_i(\vec{x})$ indicate, respectively, the overall prevalence and typicality of an attack algorithm $\mathcal{A}_i$. For simplicity, we assume flat prior $\pi_i=1/C,\; \forall i=1,\dots,C$ and focus on modeling the likelihood terms.

In practice, the true likelihood functions $P(\vec{x}|\mathcal{A}=\mathcal{A}_i)$ are replaced by a suitable class of parametric distribution 
governed by a parameter vector $\vec{\theta}_i \in \vec{\Theta}$ selected from a set $\vec{\Theta}$ with the aid of a labeled training dataset. Recall that $\vec{x}$ is a raw waveform---a random variable of varied and generally intractably high dimensionality. In our proposed probabilistic attribution framework, the key assumption is that all the relevant information for decision making can be embedded in an utterance-level embedding of fixed dimensionality. In specific, 
we make two simplifying assumptions: 
\begin{enumerate}
    \item $P_i(\vec{x})= 
P(\vec{\rho}(\vec{x})|\vec{\theta}_i)$, where $\vec{\rho}(\vec{x})$ is the probabilistic attribute embedding of $\vec{x}$; i.e. the probabilistic attribute embedding acts as a \emph{sufficient statistic} for the task of attack characterization;
\item The $L$ attribute embeddings are conditionally independent, given the attack label: 
\begin{equation}\label{eq:likelihood-score}
\begin{aligned}
    P_i(\vec{x}) &= 
    \prod_{\ell=1}^{L}P(\vec{ \rho}_\ell(\vec{x})|\vec{\theta}_i).
\end{aligned}
\end{equation}

\end{enumerate}

These assumptions lead to a simple, naive Bayes type of a classifier where we characterize each attack through empirical distributions of the attribute values. In practice, we use \emph{maximum likelihood} (ML) parameter estimator to compute the score \eqref{eq:likelihood-score}. The resulting ML estimator (computed separately per each attack) is
    \begin{equation}
        \hat{\theta}_{\ell,m}^\text{ML} = \frac{S_{\ell,m}}{\sum_{m'=1}^{M_\ell} S_{\ell,m'}},\;\;\; \ell=1,\dots,L,\;m=1,\dots,M_\ell,
    \end{equation}
where $S_{\ell,m}\mydef \sum_{n=1}^N\rho_{\ell,m}^{(n)}$ denotes the number of training observations with the value of 1 for attribute $m$ in attribute $\ell$, $N$ denoting the total number of utterances for a specific attack. The interested reader is pointed to Appendix-1 for the derivation.

\subsection{Explanability with Shapley values}



Having established a framework for probabilistic attribute-based characterization of spoofed speech and detailed the back-end classification approaches, we now turn to a crucial aspect of our methodology: explainability. While our approach allows us to classify spoofed speech accurately, understanding why a specific decision is made is equally important, especially in forensic contexts. This motivates the following use of Shapley values to quantify the contribution of each attribute, an approach that remains relatively unexplored in the speech processing literature, particularly within the context of spoofing detection and attribution. Despite their popularity in broader machine learning \cite{scott2017unified}, they have not been commonly used in speech spoofing and attribution research. Mathematically, the Shapley value of an attribute value (\(\rho_{\ell m}\)) of a probabilistic attribute embedding ($\Vec{\rho}$) using a backend classifier ($\mathcal{F}^{BC}$) is calculated as:

\begin{equation}
\begin{aligned}
\phi_{\ell m}&(\mathcal{F}^{BC},\Vec{\rho} ) =\\
                                    &\sum_{S \subseteq \Vec{\rho} \setminus {\rho_{\ell m}}} \frac{|S|!(T - |S| - 1)!}{T!} [\mathcal{F}^{BC}(S \cup {\rho_{\ell m}}) - \mathcal{F}^{BC}(S)],
\end{aligned}
\end{equation}

\noindent
where \(S\) represents a subset of features excluding \(\rho_{\ell m}\), and \(|S|\) denotes the cardinality of \(S\). The function $\mathcal{F}^{BC}$(.) denotes the prediction score output by the classifier. In practice, for any given classifier, the Shapley values are approximated by considering various combinations of the features, including and excluding (\(\rho_{\ell m}\)), to estimate the contribution of the feature to a specific prediction. The difference between these predictions, representing the marginal contribution of \(\rho_{\ell m}\), is then calculated. This process is repeated over numerous such combinations, and the Shapley value for \(\rho_{\ell m}\) is obtained by averaging these marginal contributions. In this study, we calculate the Shapley values of attributes \(\rho_{\ell m}\) to understand their contributions in performing the downstream tasks. This analysis helps us to identify the key decision-makers in the classification process and quantify their individual contributions to the downstream tasks.



\section{Experimental Setup}\label{sec:exp-setup}
In this section, we provide an 
overview of the selected datasets and detail the experimental setup, including CM embedding extraction, probabilistic attribute extraction, back-end classifiers, and the use of Shapley values~\cite{scott2017unified} for explainability. To ensure reproducibility, the implementation details and code are available on GitHub\footnote{\url{https://github.com/Manasi2001/Spoofed-Speech-Attribution}}.

\subsection{Datasets}
\label{sec:timit-db}

We use the ASVspoof2019 logical access (LA)~\cite{wang2020asvspoof} dataset for our experiments. 
The motivation for using this dataset lies in its clean, high-quality recordings, allowing us to focus deeply on explainability and interpretability in the analysis. The original dataset evaluation protocol consists of three disjoint partitions: training, development, and evaluation, each containing both bonafide (real human) and spoofed speech utterances. The spoofed utterances originate from $17$ different types of attacks, consisting of $10$ TTS and $7$ VC attacks. Six attacks are common to both the training and development sets. The evaluation set, in turn, has only $2$ common attacks (namely, A04 and A06 in the training and development sets are identical with A16 and A19 in the evaluation set) with the training and development sets, referred to as ``known" attacks. The remaining $11$ ``unknown" evaluation set attacks are disjoint from the attacks in the training and development sets. 

We utilize the dataset to address for both spoofing detection and attack attribution tasks. For the former, the original standard ASVspoof2019 protocol is used \emph{as-is}. For reasons of protocol naming consistency, and avoidance of confusion with the attribution protocols, we refer to the standard ASVspoof 2019 spoofing detection protocol as \emph{\textbf{ASVspoof2019-det}}. For the attribution task, in turn, we consider two different evaluation protocols, one that uses directly the file partitioning of the original database protocol; and another that includes repartitioning of the data in order to increase the number of attacks. In the former case, only the trials corresponding to two ``known" attacks in the evaluation set are utilized to address the closed-set attribution task, while utterances belonging to ``unknown" attacks are excluded. Hence, we refer to this two-attack attribution protocol as 
\emph{\textbf{ASVspoof2019-attr-2}}. 
Notably, the attribution protocols exclusively include spoofed utterances, excluding bonafide utterances from all three partitions: training, development, and evaluation.
To sum up, spoofed utterances in the training and development parts of ASVspoof2019-det and ASVspoof-attr-2 are identical. Except for the exclusion of bonafide data, the key difference is that the evaluation set in the latter includes data only from the two known attacks.


Since the \emph{ASVspoof2019-attr-2} protocol includes only two attacks in the evaluation set, the limitations of this protocol 
for drawing meaningful statistical conclusions are obvious. We therefore designed a new custom protocol for the attack attribution task that incorporates all the $17$ attacks from the ASVspoof2019 dataset. We refer to our protocol as the \emph{\textbf{ASVspoof2019-attr-17}} protocol and aim for it to serve as a valuable benchmark for evaluating attribution methods on the ASVspoof2019 corpus. To create this new protocol, we partitioned the original training set in an 80:20 ratio to derive the new training and development sets for A01-A06 attacks, with the original development set repurposed as the evaluation set for these attacks. For A07-A19 attacks, the original evaluation set was divided in a 50:10:40 ratio for each attack to create the training, development, and evaluation partitions, with the evaluation set comprising $12,948$ utterances from the speaker-common set ($39$ speakers) and $12,636$ utterances from the speaker-disjoint set ($9$ speakers).  Utterances in the speaker-common set are shared across the training, development, and evaluation partitions, whereas utterances in the speaker-disjoint set are included only in the evaluation partition. To enhance speaker-independent training for these attacks, a higher number of speakers were allocated to the speaker-common set compared to the speaker-disjoint set. Speaker identities were extracted from the generation protocol released by the ASVspoof consortium\footnote{\url{https://www.asvspoof.org/asvspoof2019/ASVspoof2019_2021_VCTK_VCC_MetaInfo.tar.gz}}. This newly formed attribution protocol now includes all $17$ attacks across the training, development, and evaluation partitions. The utterance statistics for the ASVspoof2019-det, ASVspoof2019-attr-2, and the newly created ASVspoof2019-attr-17 protocols are provided in Table~\ref{tab:ASVspoof2019-att-stat}.

 Although a customized attribution protocol for ASVspoof2019 was previously used in ~\cite{borrelli2021synthetic}, it was not made publicly available, unlike our proposed protocol\footnote{\url{https://github.com/jagabandhumishra/ASVspoof2019-Attack-Attribution-Protocol}}. Furthermore, In ~\cite{zhu2022source}\footnote{\url{https://github.com/kurisujhin/anti-spoof-source-tracing}}, while the protocol for source tracing was made public, the task primarily focused on classifying the attribute values of three attributes—conversion, speaker representation, and waveform generation—out of the seven available in the ASVspoof2019 metadata. In contrast, our approach aims to classify the attribute values of all seven attributes, leveraging them to extract probabilistic attribute embeddings and subsequently using these embeddings for spoofing attack attribution. Notably, spoofing attacks are defined by specific combinations of attribute values. However, the referenced protocol introduces inconsistencies in spoofing attacks across partitions due to its focus on classifying the attribute values of only three attributes. Specifically, the training partition includes attacks A02–A04, A06, and A08–A19, while the evaluation partition includes attacks A01, A05, and A07. This inconsistency in attack partitions makes the protocol unsuitable for spoofing attack attribution, which requires consistent treatment of spoofing attacks across both training and evaluation sets.

\begin{table}[]
\centering
\caption{ASVspoof2019-det, ASVspoof2019-attr-2$^{*}$ and ASVspoof2019-attr-17, utterances per attack statistics. F and M correspond to females and males, respectively. \#Spk and \#Utter: correspond to the number of speakers and utterances, respectively. Blank space signifies no utterance. The same color between ASVspoof2019 and ASVspoof2019 attack attribution signifies the use of the same set of utterances. $^{*}$ The evaluation set of the ASVspoof2019-attr-2 protocol used for the spoofing attack attribution task consists only of known attacks A16 and A19. Furthermore, both the ASVspoof2019-attr-2 and ASVspoof2019-attr-17 protocols contain only spoofed utterances and do not include any bonafide speech. In contrast, bonafide utterances are present only in the ASVspoof2019-det protocol, designed for spoofing detection (i.e. bonafide/spoof) classification}.
\label{tab:ASVspoof2019-att-stat}
\resizebox{\columnwidth}{!}{%
\begin{tabular}{|c|cccccc|cccccc|}
\hline
 &
  \multicolumn{6}{c|}{} &
  \multicolumn{6}{c|}{} \\
 &
  \multicolumn{6}{c|}{\multirow{-2}{*}{ASVspoof2019-det and ASVspoof2019-attr-2}} &
  \multicolumn{6}{c|}{\multirow{-2}{*}{ASVspoof2019-attr-17}} \\ \cline{2-13} 
 &
  \multicolumn{2}{c|}{Train} &
  \multicolumn{2}{c|}{Dev} &
  \multicolumn{2}{c|}{Eval} &
  \multicolumn{2}{c|}{Train} &
  \multicolumn{2}{c|}{Dev} &
  \multicolumn{2}{c|}{Eval} \\ \cline{2-13} 
\multirow{-4}{*}{} &
  \multicolumn{1}{c|}{\#Utter} &
  \multicolumn{1}{c|}{\#Spk} &
  \multicolumn{1}{c|}{\#Utter} &
  \multicolumn{1}{c|}{\#Spk} &
  \multicolumn{1}{c|}{\#Utter} &
  \#Spk &
  \multicolumn{1}{c|}{\#Utter} &
  \multicolumn{1}{c|}{\#Spk} &
  \multicolumn{1}{c|}{\#Utter} &
  \multicolumn{1}{c|}{\#Spk} &
  \multicolumn{1}{c|}{\#Utter} &
  \#Spk \\ \hline
\rowcolor[HTML]{D9D9D9} 
Bonafide &
  \multicolumn{1}{c|}{\cellcolor[HTML]{D9D9D9}2580} &
  \multicolumn{1}{l|}{\cellcolor[HTML]{D9D9D9}12F,8M} &
  \multicolumn{1}{c|}{\cellcolor[HTML]{D9D9D9}2548} &
  \multicolumn{1}{c|}{\cellcolor[HTML]{D9D9D9}12M,8F} &
  \multicolumn{1}{c|}{\cellcolor[HTML]{D9D9D9}7355} &
  37F,30M &
  \multicolumn{6}{c|}{\cellcolor[HTML]{D9D9D9}} \\ \hline
A01 &
  \multicolumn{1}{c|}{\cellcolor[HTML]{C0F1C8}3800} &
  \multicolumn{1}{c|}{\cellcolor[HTML]{C0F1C8}} &
  \multicolumn{1}{c|}{\cellcolor[HTML]{FBE2D5}3716} &
  \multicolumn{1}{c|}{\cellcolor[HTML]{FBE2D5}} &
  \multicolumn{2}{c|}{\cellcolor[HTML]{CAEDFB}} &
  \multicolumn{1}{c|}{\cellcolor[HTML]{C0F1C8}3040} &
  \multicolumn{1}{c|}{\cellcolor[HTML]{C0F1C8}} &
  \multicolumn{1}{c|}{\cellcolor[HTML]{C0F1C8}760} &
  \multicolumn{1}{c|}{\cellcolor[HTML]{C0F1C8}} &
  \multicolumn{1}{c|}{\cellcolor[HTML]{FBE2D5}3716} &
  \cellcolor[HTML]{FBE2D5} \\ \cline{1-2} \cline{4-4} \cline{8-8} \cline{10-10} \cline{12-12}
A02 &
  \multicolumn{1}{c|}{\cellcolor[HTML]{C0F1C8}3800} &
  \multicolumn{1}{c|}{\cellcolor[HTML]{C0F1C8}} &
  \multicolumn{1}{c|}{\cellcolor[HTML]{FBE2D5}3716} &
  \multicolumn{1}{c|}{\cellcolor[HTML]{FBE2D5}} &
  \multicolumn{2}{c|}{\cellcolor[HTML]{CAEDFB}} &
  \multicolumn{1}{c|}{\cellcolor[HTML]{C0F1C8}3040} &
  \multicolumn{1}{c|}{\cellcolor[HTML]{C0F1C8}} &
  \multicolumn{1}{c|}{\cellcolor[HTML]{C0F1C8}760} &
  \multicolumn{1}{c|}{\cellcolor[HTML]{C0F1C8}} &
  \multicolumn{1}{c|}{\cellcolor[HTML]{FBE2D5}3716} &
  \cellcolor[HTML]{FBE2D5} \\ \cline{1-2} \cline{4-4} \cline{8-8} \cline{10-10} \cline{12-12}
A03 &
  \multicolumn{1}{c|}{\cellcolor[HTML]{C0F1C8}3800} &
  \multicolumn{1}{c|}{\cellcolor[HTML]{C0F1C8}} &
  \multicolumn{1}{c|}{\cellcolor[HTML]{FBE2D5}3716} &
  \multicolumn{1}{c|}{\cellcolor[HTML]{FBE2D5}} &
  \multicolumn{2}{c|}{\cellcolor[HTML]{CAEDFB}} &
  \multicolumn{1}{c|}{\cellcolor[HTML]{C0F1C8}3040} &
  \multicolumn{1}{c|}{\cellcolor[HTML]{C0F1C8}} &
  \multicolumn{1}{c|}{\cellcolor[HTML]{C0F1C8}760} &
  \multicolumn{1}{c|}{\cellcolor[HTML]{C0F1C8}} &
  \multicolumn{1}{c|}{\cellcolor[HTML]{FBE2D5}3716} &
  \cellcolor[HTML]{FBE2D5} \\ \cline{1-2} \cline{4-4} \cline{8-8} \cline{10-10} \cline{12-12}
A04$^{*}$ &
  \multicolumn{1}{c|}{\cellcolor[HTML]{C0F1C8}3800} &
  \multicolumn{1}{c|}{\cellcolor[HTML]{C0F1C8}} &
  \multicolumn{1}{c|}{\cellcolor[HTML]{FBE2D5}3716} &
  \multicolumn{1}{c|}{\cellcolor[HTML]{FBE2D5}} &
  \multicolumn{2}{c|}{\cellcolor[HTML]{CAEDFB}} &
  \multicolumn{1}{c|}{\cellcolor[HTML]{C0F1C8}3040} &
  \multicolumn{1}{c|}{\cellcolor[HTML]{C0F1C8}} &
  \multicolumn{1}{c|}{\cellcolor[HTML]{C0F1C8}760} &
  \multicolumn{1}{c|}{\cellcolor[HTML]{C0F1C8}} &
  \multicolumn{1}{c|}{\cellcolor[HTML]{FBE2D5}3716} &
  \cellcolor[HTML]{FBE2D5} \\ \cline{1-2} \cline{4-4} \cline{8-8} \cline{10-10} \cline{12-12}
A05 &
  \multicolumn{1}{c|}{\cellcolor[HTML]{C0F1C8}3800} &
  \multicolumn{1}{c|}{\cellcolor[HTML]{C0F1C8}} &
  \multicolumn{1}{c|}{\cellcolor[HTML]{FBE2D5}3716} &
  \multicolumn{1}{c|}{\cellcolor[HTML]{FBE2D5}} &
  \multicolumn{2}{c|}{\cellcolor[HTML]{CAEDFB}} &
  \multicolumn{1}{c|}{\cellcolor[HTML]{C0F1C8}3040} &
  \multicolumn{1}{c|}{\cellcolor[HTML]{C0F1C8}} &
  \multicolumn{1}{c|}{\cellcolor[HTML]{C0F1C8}760} &
  \multicolumn{1}{c|}{\cellcolor[HTML]{C0F1C8}} &
  \multicolumn{1}{c|}{\cellcolor[HTML]{FBE2D5}3716} &
  \cellcolor[HTML]{FBE2D5} \\ \cline{1-2} \cline{4-4} \cline{8-8} \cline{10-10} \cline{12-12}
A06$^{*}$ &
  \multicolumn{1}{c|}{\cellcolor[HTML]{C0F1C8}3800} &
  \multicolumn{1}{c|}{\multirow{-6}{*}{\cellcolor[HTML]{C0F1C8}12F,8M}} &
  \multicolumn{1}{c|}{\cellcolor[HTML]{FBE2D5}3716} &
  \multicolumn{1}{c|}{\multirow{-6}{*}{\cellcolor[HTML]{FBE2D5}6F,4M}} &
  \multicolumn{2}{c|}{\multirow{-6}{*}{\cellcolor[HTML]{CAEDFB}}} &
  \multicolumn{1}{c|}{\cellcolor[HTML]{C0F1C8}3040} &
  \multicolumn{1}{c|}{\multirow{-6}{*}{\cellcolor[HTML]{C0F1C8}12F,8M}} &
  \multicolumn{1}{c|}{\cellcolor[HTML]{C0F1C8}760} &
  \multicolumn{1}{c|}{\multirow{-6}{*}{\cellcolor[HTML]{C0F1C8}12F,8M}} &
  \multicolumn{1}{c|}{\cellcolor[HTML]{FBE2D5}3716} &
  \multirow{-6}{*}{\cellcolor[HTML]{FBE2D5}6F,4M} \\ \hline
A07 &
  \multicolumn{2}{c|}{\cellcolor[HTML]{CAEDFB}} &
  \multicolumn{2}{c|}{\cellcolor[HTML]{CAEDFB}} &
  \multicolumn{1}{c|}{\cellcolor[HTML]{82CAEC}4914} &
  \cellcolor[HTML]{82CAEC} &
  \multicolumn{1}{c|}{\cellcolor[HTML]{82CAEC}2357} &
  \multicolumn{1}{c|}{\cellcolor[HTML]{82CAEC}} &
  \multicolumn{1}{c|}{\cellcolor[HTML]{82CAEC}589} &
  \multicolumn{1}{c|}{\cellcolor[HTML]{82CAEC}} &
  \multicolumn{1}{c|}{\cellcolor[HTML]{82CAEC}1968} &
  \cellcolor[HTML]{82CAEC} \\ \cline{1-1} \cline{6-6} \cline{8-8} \cline{10-10} \cline{12-12}
A08 &
  \multicolumn{2}{c|}{\cellcolor[HTML]{CAEDFB}} &
  \multicolumn{2}{c|}{\cellcolor[HTML]{CAEDFB}} &
  \multicolumn{1}{c|}{\cellcolor[HTML]{82CAEC}4914} &
  \cellcolor[HTML]{82CAEC} &
  \multicolumn{1}{c|}{\cellcolor[HTML]{82CAEC}2357} &
  \multicolumn{1}{c|}{\cellcolor[HTML]{82CAEC}} &
  \multicolumn{1}{c|}{\cellcolor[HTML]{82CAEC}589} &
  \multicolumn{1}{c|}{\cellcolor[HTML]{82CAEC}} &
  \multicolumn{1}{c|}{\cellcolor[HTML]{82CAEC}1968} &
  \cellcolor[HTML]{82CAEC} \\ \cline{1-1} \cline{6-6} \cline{8-8} \cline{10-10} \cline{12-12}
A09 &
  \multicolumn{2}{c|}{\cellcolor[HTML]{CAEDFB}} &
  \multicolumn{2}{c|}{\cellcolor[HTML]{CAEDFB}} &
  \multicolumn{1}{c|}{\cellcolor[HTML]{82CAEC}4914} &
  \cellcolor[HTML]{82CAEC} &
  \multicolumn{1}{c|}{\cellcolor[HTML]{82CAEC}2357} &
  \multicolumn{1}{c|}{\cellcolor[HTML]{82CAEC}} &
  \multicolumn{1}{c|}{\cellcolor[HTML]{82CAEC}589} &
  \multicolumn{1}{c|}{\cellcolor[HTML]{82CAEC}} &
  \multicolumn{1}{c|}{\cellcolor[HTML]{82CAEC}1968} &
  \cellcolor[HTML]{82CAEC} \\ \cline{1-1} \cline{6-6} \cline{8-8} \cline{10-10} \cline{12-12}
A10 &
  \multicolumn{2}{c|}{\cellcolor[HTML]{CAEDFB}} &
  \multicolumn{2}{c|}{\cellcolor[HTML]{CAEDFB}} &
  \multicolumn{1}{c|}{\cellcolor[HTML]{82CAEC}4914} &
  \cellcolor[HTML]{82CAEC} &
  \multicolumn{1}{c|}{\cellcolor[HTML]{82CAEC}2357} &
  \multicolumn{1}{c|}{\cellcolor[HTML]{82CAEC}} &
  \multicolumn{1}{c|}{\cellcolor[HTML]{82CAEC}589} &
  \multicolumn{1}{c|}{\cellcolor[HTML]{82CAEC}} &
  \multicolumn{1}{c|}{\cellcolor[HTML]{82CAEC}1968} &
  \cellcolor[HTML]{82CAEC} \\ \cline{1-1} \cline{6-6} \cline{8-8} \cline{10-10} \cline{12-12}
A11 &
  \multicolumn{2}{c|}{\cellcolor[HTML]{CAEDFB}} &
  \multicolumn{2}{c|}{\cellcolor[HTML]{CAEDFB}} &
  \multicolumn{1}{c|}{\cellcolor[HTML]{82CAEC}4914} &
  \cellcolor[HTML]{82CAEC} &
  \multicolumn{1}{c|}{\cellcolor[HTML]{82CAEC}2356} &
  \multicolumn{1}{c|}{\cellcolor[HTML]{82CAEC}} &
  \multicolumn{1}{c|}{\cellcolor[HTML]{82CAEC}590} &
  \multicolumn{1}{c|}{\cellcolor[HTML]{82CAEC}} &
  \multicolumn{1}{c|}{\cellcolor[HTML]{82CAEC}1968} &
  \cellcolor[HTML]{82CAEC} \\ \cline{1-1} \cline{6-6} \cline{8-8} \cline{10-10} \cline{12-12}
A12 &
  \multicolumn{2}{c|}{\cellcolor[HTML]{CAEDFB}} &
  \multicolumn{2}{c|}{\cellcolor[HTML]{CAEDFB}} &
  \multicolumn{1}{c|}{\cellcolor[HTML]{82CAEC}4914} &
  \cellcolor[HTML]{82CAEC} &
  \multicolumn{1}{c|}{\cellcolor[HTML]{82CAEC}2357} &
  \multicolumn{1}{c|}{\cellcolor[HTML]{82CAEC}} &
  \multicolumn{1}{c|}{\cellcolor[HTML]{82CAEC}589} &
  \multicolumn{1}{c|}{\cellcolor[HTML]{82CAEC}} &
  \multicolumn{1}{c|}{\cellcolor[HTML]{82CAEC}1968} &
  \cellcolor[HTML]{82CAEC} \\ \cline{1-1} \cline{6-6} \cline{8-8} \cline{10-10} \cline{12-12}
A13 &
  \multicolumn{2}{c|}{\cellcolor[HTML]{CAEDFB}} &
  \multicolumn{2}{c|}{\cellcolor[HTML]{CAEDFB}} &
  \multicolumn{1}{c|}{\cellcolor[HTML]{82CAEC}4914} &
  \cellcolor[HTML]{82CAEC} &
  \multicolumn{1}{c|}{\cellcolor[HTML]{82CAEC}2356} &
  \multicolumn{1}{c|}{\cellcolor[HTML]{82CAEC}} &
  \multicolumn{1}{c|}{\cellcolor[HTML]{82CAEC}590} &
  \multicolumn{1}{c|}{\cellcolor[HTML]{82CAEC}} &
  \multicolumn{1}{c|}{\cellcolor[HTML]{82CAEC}1968} &
  \cellcolor[HTML]{82CAEC} \\ \cline{1-1} \cline{6-6} \cline{8-8} \cline{10-10} \cline{12-12}
A14 &
  \multicolumn{2}{c|}{\cellcolor[HTML]{CAEDFB}} &
  \multicolumn{2}{c|}{\cellcolor[HTML]{CAEDFB}} &
  \multicolumn{1}{c|}{\cellcolor[HTML]{82CAEC}4914} &
  \cellcolor[HTML]{82CAEC} &
  \multicolumn{1}{c|}{\cellcolor[HTML]{82CAEC}2357} &
  \multicolumn{1}{c|}{\cellcolor[HTML]{82CAEC}} &
  \multicolumn{1}{c|}{\cellcolor[HTML]{82CAEC}589} &
  \multicolumn{1}{c|}{\cellcolor[HTML]{82CAEC}} &
  \multicolumn{1}{c|}{\cellcolor[HTML]{82CAEC}1968} &
  \cellcolor[HTML]{82CAEC} \\ \cline{1-1} \cline{6-6} \cline{8-8} \cline{10-10} \cline{12-12}
A15 &
  \multicolumn{2}{c|}{\cellcolor[HTML]{CAEDFB}} &
  \multicolumn{2}{c|}{\cellcolor[HTML]{CAEDFB}} &
  \multicolumn{1}{c|}{\cellcolor[HTML]{82CAEC}4914} &
  \cellcolor[HTML]{82CAEC} &
  \multicolumn{1}{c|}{\cellcolor[HTML]{82CAEC}2357} &
  \multicolumn{1}{c|}{\cellcolor[HTML]{82CAEC}} &
  \multicolumn{1}{c|}{\cellcolor[HTML]{82CAEC}589} &
  \multicolumn{1}{c|}{\cellcolor[HTML]{82CAEC}} &
  \multicolumn{1}{c|}{\cellcolor[HTML]{82CAEC}1968} &
  \cellcolor[HTML]{82CAEC} \\ \cline{1-1} \cline{6-6} \cline{8-8} \cline{10-10} \cline{12-12}
A16 (A04)$^{*}$ &
  \multicolumn{2}{c|}{\cellcolor[HTML]{CAEDFB}} &
  \multicolumn{2}{c|}{\cellcolor[HTML]{CAEDFB}} &
  \multicolumn{1}{c|}{\cellcolor[HTML]{82CAEC}4914} &
  \cellcolor[HTML]{82CAEC} &
  \multicolumn{1}{c|}{\cellcolor[HTML]{82CAEC}2357} &
  \multicolumn{1}{c|}{\cellcolor[HTML]{82CAEC}} &
  \multicolumn{1}{c|}{\cellcolor[HTML]{82CAEC}589} &
  \multicolumn{1}{c|}{\cellcolor[HTML]{82CAEC}} &
  \multicolumn{1}{c|}{\cellcolor[HTML]{82CAEC}1968} &
  \cellcolor[HTML]{82CAEC} \\ \cline{1-1} \cline{6-6} \cline{8-8} \cline{10-10} \cline{12-12}
A17 &
  \multicolumn{2}{c|}{\cellcolor[HTML]{CAEDFB}} &
  \multicolumn{2}{c|}{\cellcolor[HTML]{CAEDFB}} &
  \multicolumn{1}{c|}{\cellcolor[HTML]{82CAEC}4914} &
  \cellcolor[HTML]{82CAEC} &
  \multicolumn{1}{c|}{\cellcolor[HTML]{82CAEC}2357} &
  \multicolumn{1}{c|}{\cellcolor[HTML]{82CAEC}} &
  \multicolumn{1}{c|}{\cellcolor[HTML]{82CAEC}589} &
  \multicolumn{1}{c|}{\cellcolor[HTML]{82CAEC}} &
  \multicolumn{1}{c|}{\cellcolor[HTML]{82CAEC}1968} &
  \cellcolor[HTML]{82CAEC} \\ \cline{1-1} \cline{6-6} \cline{8-8} \cline{10-10} \cline{12-12}
A18 &
  \multicolumn{2}{c|}{\cellcolor[HTML]{CAEDFB}} &
  \multicolumn{2}{c|}{\cellcolor[HTML]{CAEDFB}} &
  \multicolumn{1}{c|}{\cellcolor[HTML]{82CAEC}4914} &
  \cellcolor[HTML]{82CAEC} &
  \multicolumn{1}{c|}{\cellcolor[HTML]{82CAEC}2357} &
  \multicolumn{1}{c|}{\cellcolor[HTML]{82CAEC}} &
  \multicolumn{1}{c|}{\cellcolor[HTML]{82CAEC}589} &
  \multicolumn{1}{c|}{\cellcolor[HTML]{82CAEC}} &
  \multicolumn{1}{c|}{\cellcolor[HTML]{82CAEC}1968} &
  \cellcolor[HTML]{82CAEC} \\ \cline{1-1} \cline{6-6} \cline{8-8} \cline{10-10} \cline{12-12}
A19 (A06)$^{*}$ &
  \multicolumn{2}{c|}{\multirow{-13}{*}{\cellcolor[HTML]{CAEDFB}}} &
  \multicolumn{2}{c|}{\multirow{-13}{*}{\cellcolor[HTML]{CAEDFB}}} &
  \multicolumn{1}{c|}{\cellcolor[HTML]{82CAEC}4914} &
  \multirow{-13}{*}{\cellcolor[HTML]{82CAEC}27F,21M} &
  \multicolumn{1}{c|}{\cellcolor[HTML]{82CAEC}2356} &
  \multicolumn{1}{c|}{\multirow{-13}{*}{\cellcolor[HTML]{82CAEC}21F,18M}} &
  \multicolumn{1}{c|}{\cellcolor[HTML]{82CAEC}590} &
  \multicolumn{1}{c|}{\multirow{-13}{*}{\cellcolor[HTML]{82CAEC}21F,18M}} &
  \multicolumn{1}{c|}{\cellcolor[HTML]{82CAEC}1968} &
  \multirow{-13}{*}{\cellcolor[HTML]{82CAEC}\begin{tabular}[c]{@{}c@{}} Speaker Common\\ (21F,18M),\\  Speaker Disjoint\\ (6F,3M)\end{tabular}} \\ \hline
Total (\#Spoof) &
  \multicolumn{2}{c|}{22800} &
  \multicolumn{2}{c|}{22296} &
  \multicolumn{2}{c|}{63882} &
  \multicolumn{2}{c|}{48878} &
  \multicolumn{2}{c|}{12220} &
  \multicolumn{2}{c|}{47880} \\ \hline
Total (\#Spoof) &
  \multicolumn{6}{c|}{108978} &
  \multicolumn{6}{c|}{108978} \\ \hline
\end{tabular}%
}
\end{table}
\subsection{CM embedding extraction}

As described above, the presented attribution methodology builds upon recording-level embeddings $\vec{e}_\text{cm}$ extracted through a spoofing countermeasure (CM) system. In this work, we consider 
three well-established CMs: \emph{AASIST}~\cite{jung2022aasist}, \emph{Rawboost} AASIST ~\cite{tak2022rawboost}, and \emph{self-supervised} AASIST (SSL-AASIST)~\cite{tak2022automatic}.
All three are based on AASIST~\cite{jung2022aasist}, a widely used CM system that operates directly on the raw waveform. It consists of a RawNet2~\cite{jung2020improved} 
encoder and two graph modules, graph attention and graph pooling layers, on top of the encoder. It allows feature extraction considering both temporal and spectral dimensions. The
graph modules first separately model spectral and temporal features, which are then combined 
to a heterogeneous graph. Specifically, two heterogeneous graphs are generated to extract prominent features in a competitive manner, similar to \emph{max feature map} used in ~\cite{wu2018lcnn}.

Considering model generalization, we employ two variations of the AASIST model, RawBoost~\cite{tak2022rawboost} and an SSL-based model~\cite{tak2022automatic}, applied on top of the vanilla AASIST model.  Although AASIST-RawBoost is typically associated with improving generalization of spoofing detection to degraded conditions, we 
include it along with vanilla and SSL variation of AASIST for contrastive purposes, given they have received less attention in ’clean’ conditions and in attribution tasks.

The RawBoost-AASIST includes three different data augmentation techniques: (i) linear and non-linear convolutive noise, (ii) impulsive signal-dependent additive noise, and (iii) stationary signal-independent additive noise. Linear and nonlinear convolutive noise is introduced to the signal to simulate the effects of encoding, decoding, and compression-decompression processes. Impulsive signal-dependent additive noise mimics the noise caused by improper microphone placement and quantization errors, while stationary signal-independent additive noise simulates the noise introduced by loose/poorly connected cables, air conditioning systems, or other electronic sources. All three data augmentation techniques are applied concurrently in our approach. The second model variant, SSL-AASIST, uses both a pre-trained wav2vec 2.0 model \cite{baevski2020wav2vec} (trained in a self-supervised manner) and data augmentation for improved generalization to different domains. It replaces the sinc-layer front-end components with the wav2vec 2.0 model. The model is fine-tuned using in-domain ASVspoof2019 data as pre-training is only conducted over bonafide data.



Using the 
embeddings ($\vec{e}_\text{cm}$) extracted through the three CMs 
we 
train the probabilistic attribute extractors 
$\mathcal{F}^{\text{pae}}_{{\bm \rho}_{l}}$ used for extracting the probabilistic attribute embeddings (${\bm \rho}$). 
The overall pipeline of the proposed explainable architecture is depicted in Figure~\ref{Blk_diag_overall}, with details of the probabilistic attribute extractor being discussed in the following sub-section.

\subsection{Probabilistic Attribute Extraction}

We train the probabilistic attribute extractors $\{\mathcal{F}^{\text{pae}}_{{\bm \rho}_{\ell}}\}$ for each attribute 
separately, each one 
consisting of three fully-connected layers. The first two layers have $64$ and $32$ neurons, while the final output layer contains $M_{\ell}$ neurons, corresponding to the possible values of attribute ${\ell}$. The first two layers have rectified linear units (ReLu) and output layer neurons use a softmax activation function. Each attribute extractor is trained for $100$ epochs using the Adam optimizer~\cite{kingma2014adam} with a learning rate of $0.0001$.  The architecture and hyperparameters were decided after running several experiments on the original ASVspoof2019 protocol and evaluating the performance of probabilistic attribute extraction on the development set. It is our intentional choice to fix 
the same network architecture (apart from the output layer) and use identical training recipe for all the attributes. This facilitates more commensurable comparison of the attribute extractor performance easier. 

The training portion of the dataset is used to train the attribute extractors, and the development portion serves for model selection.
The epoch that corresponds to the lowest equal error rate (EER) for detecting attribute values 
in the development set is selected for probabilistic attribute embedding extraction. Since training involves generally a multi-class setup, but EER is a discrimination measure between \emph{two} classes only, we consider the value that is actually present in a particular development utterance as a target attribute value (positive class), treating the remaining $M_{\ell} - 1$ values as non-target values (negative class). 
The distribution of all the target and non-target development scores is then used to calculate the EER. 

Let us now detail the attribute-value pairs used in developing the attribute extractors. 
All ASVspoof 2019 
utterances 
were generated using a modular approach. We use the respective 
metadata 
available in \cite[Table~1]{wang2020asvspoof} 
to design and train seven 
extractors following the attribute-value pairs, as summarized in Table~\ref{tab:att-val-asvspoof}. 
The spoofed utterances including both VC and TTS are generated based on predefined high-level design choices, which include input, input processor, duration model, conversion function, speaker representation, conversion model outputs, and waveform generator (voice coder or vocoder). 
Note that attribute-value pair characterization depends solely on spoofed speech generation and is independent of bonafide speech. Additionally, both ASVspoof2019-det and ASVspoof2019-attr-2 protocols have identical spoofing attacks in the training and development partitions---hence, the derived attribute and value pairs for both protocols are identical.  
The 
ASVspoof2019-attr-17 protocol, in turn, 
contains all the $17$ attacks 
(note that attack pairs (A04, A16) and  (A06, A19) 
present identical methods). Compared 
with the two other evaluation protocols, 
the attribute definitions are identical, but their value sets 
are now expanded due to the inclusion of more 
attacks. 
To sum up, the attribute values are dictated by the metadata available. Note also that the dimensionality of the probabilistic embeddings in ASVspoof2019-attr-17 is twice of that in the two other protocols. 

\begin{table}[]
\centering
\caption{Derived attributes and their values from ASVspoof2019-det, ASVspoof2019-attr-2 and ASVspoof2019-attr-17 protocols. The training and development partition of ASVspoof2019-det and ASVspoof2019-attr-2 are identical, hence the attribute and value pairs are identical for both protocols. \#attVal signifies the number of attribute values. }
\vspace{0.2 cm}
\label{tab:att-val-asvspoof}
\resizebox{\columnwidth}{!}{%
\begin{tabular}{|c|cc|cc|}
\hline
\multirow{2}{*}{Attributes} &
  \multicolumn{2}{c|}{ASVspoof2019-det and ASVspoof2019-attr-2} &
  \multicolumn{2}{c|}{ASVspoof2019-attr-17} \\ \cline{2-5} 
 &
  \multicolumn{1}{c|}{Values} &
  \#attVal &
  \multicolumn{1}{c|}{Values} &
  \#attVal \\ \hline
\begin{tabular}[c]{@{}c@{}}Attribute 1 \\ (inputs)\end{tabular} &
  \multicolumn{1}{c|}{\{Text, Speech (human)\}} &
  2 &
  \multicolumn{1}{c|}{\{Text, Speech (human), Speech (TTS)\}} &
  3 \\ \hline
\begin{tabular}[c]{@{}c@{}}Attribute 2 \\ (input processor)\end{tabular} &
  \multicolumn{1}{c|}{\{NLP, WORLD, LPCC/MFCC\}} &
  3 &
  \multicolumn{1}{c|}{\begin{tabular}[c]{@{}c@{}}\{NLP, WORLD, LPCC/MFCC, CNN+bi-RNN, \\ ASR, MFCC/i-vector\}\end{tabular}} &
  6 \\ \hline
\begin{tabular}[c]{@{}c@{}}Attribute 3 \\ (duration)\end{tabular} &
  \multicolumn{1}{c|}{\{HMM, FF, None\}} &
  3 &
  \multicolumn{1}{c|}{\{HMM, FF, RNN, Attention, DTW, none\}} &
  6 \\ \hline
\begin{tabular}[c]{@{}c@{}}Attribute 4 \\ (conversion)\end{tabular} &
  \multicolumn{1}{c|}{\{AR-RNN, FF, CART, VAE, GMM-UBM\}} &
  5 &
  \multicolumn{1}{c|}{\begin{tabular}[c]{@{}c@{}}\{AR-RNN, FF, CART, VAE, GMM-UBM, RNN, \\ AR-RNN+CNN, Moment-match, linear\}\end{tabular}} &
  9 \\ \hline
\begin{tabular}[c]{@{}c@{}}Attribute 5 \\ (speaker)\end{tabular} &
  \multicolumn{1}{c|}{\{VAE, one-hot, None\}} &
  3 &
  \multicolumn{1}{c|}{\{VAE, one-hot, d-vector (RNN), PLDA, none\}} &
  5 \\ \hline
\begin{tabular}[c]{@{}c@{}}Attribute 6 \\ (outputs)\end{tabular} &
  \multicolumn{1}{c|}{\{MCC-F0, MFCC-F0-BAP, MFCC-F0, MFCC-F0-AP, LPC\}} &
  5 &
  \multicolumn{1}{c|}{\begin{tabular}[c]{@{}c@{}}\{MCC-F0, MFCC-F0-BAP, MFCC-F0, MFCC-F0-AP, LPC, \\ MCC-F0-BA, mel-spec, F0+ling, MCC, MFCC\}\end{tabular}} &
  10 \\ \hline
\begin{tabular}[c]{@{}c@{}}Attribute 7 \\ (waveform gen)\end{tabular} &
  \multicolumn{1}{c|}{\{WaveNet, WORLD, Concat, SpecFiltOLA\}} &
  4 &
  \multicolumn{1}{c|}{\begin{tabular}[c]{@{}c@{}}\{WaveNet, WORLD, Concat, SpecFiltOLA, NeuralFilt, Vocaine, \\ WaveRNN, GriffinLim, WaveFilt, STRAIGHT, MFCCvoc\}\end{tabular}} &
  11 \\ \hline
Total \#attVal &
  \multicolumn{2}{c|}{\textbf{25}} &
  \multicolumn{2}{c|}{\textbf{50}} \\ \hline
\end{tabular}%
}
\end{table}

\subsection{Back-end Classifiers}

Up to this point, we have described the extraction of recording-level embeddings, whether the raw CM embedding ($\vec{e}_\text{cm}$), or the proposed probabilistic attribute embedding ($\vec{\rho}$) extracted through multilayer perceptrons (MLPs). The former acts as information bottleneck for the latter, intended to add transparency to decisions making by the back-end classifier. 

We consider four 
alternative back-end classifiers. The first one is the \textbf{naive Bayes (NB)}~\cite{bishop2006pattern} approach uses empirical distributions to represent attack-specific distributions of the probabilistic attributes, as detailed in Section \ref{section:downstram-task}. Second, we include a classic \textbf{decision tree (DT)} model~\cite{breiman2017classification}, a simple nonlinear approach that is particularly well-fitted for the explainability perspective. The final two back-end classifiers used are \textbf{logistic regression (LR)}~\cite{mccullagh2019generalized} and \textbf{support vector machine (SVM)} with a linear kernel~\cite{patle2013svm}, both 
well-established linear classifiers known for their practical 
generalization capability. Recall that our downstream tasks include spoofing detection, a binary classification (detection) task, and spoofing attack attribution, a multiclass task. Since LR and SVM were originally designed for the former case, 
for the 
multiclass use we adopt the one-vs-rest strategy~\cite{bishop2006pattern}, 
which involves training a separate binary classifier for each class, treating that class as the target (positive) class, with all the other classes treated as non-target (negative) classes. The classification decision is 
obtained by selecting the class with the highest confidence score among all the individual binary classifiers.




In practice, we use \texttt{scikit-learn} libraries\footnote{\url{https://scikit-learn.org/stable/}} to train and evaluate decision trees, logistic regression, and SVM classifiers. For the ASVspoof2019-det and ASVspoof2019-attr-2, the maximum depth of the decision trees is set to $5$, while for the ASVspoof2019-attr-17 the max-depth parameter is set to $15$. These parameters were fixed by observing the 
performance on the development set. 
A reference implementation of the 
NB classifier is available at\footnote{\url{https://github.com/Manasi2001/Spoofed-Speech-Attribution}}. In all the experiments, the back-end classifiers are trained in the training set of the corresponding dataset and evaluated in the evaluation set of the corresponding dataset.

\subsection{Performance metrics}
The performance of spoofing detection and spoofing attack attribution is evaluated using balanced accuracy\footnote{\url{https://scikit-learn.org/dev/modules/model_evaluation.html#balanced-accuracy-score}} and equal error rate (EER). 
Balanced accuracy is used instead of standard accuracy because the evaluation set for both tasks exhibits a skewed class-specific utterance distribution. It is calculated as follows:


\begin{equation}
    \text{Balanced accuracy}=\frac{1}{C}\sum_{i=1}^{C}\frac{\text{TP(i)}}{\text{TP(i)}+\text{FN(i)}}~,
\end{equation}
where $C$,
\(\text{TP(i)}\) and \(\text{FN(i)}\) denote, respectively, the total number of classes, the number of true positives and false negatives for class $i$. The former 
is the number of class \(i\) utterances that are predicted as class \(i\) 
and the latter is the number of class \(i\) utterances that are not predicted as class \(i\). 


\subsection{Explanability with Shapley values}

To gain deeper insights into the decision-making processes of our proposed approach, we conducted a Shapley value~\cite{scott2017unified} analysis to quantify the contribution of each feature (i.e., each attribute value) to the predictions. This analysis was performed using the best-performing back-end classifier for each downstream task. Shapley values were calculated on a per-sample basis, meaning that the contribution of each feature was specific to a particular speech utterance in the evaluation set. Since Shapley values can be positive (indicating a positive influence on the decision) or negative (indicating a negative influence), both directions contribute meaningfully to the final prediction. Therefore, we consider the absolute Shapley values in our study to measure overall influence, regardless of direction. The contribution of attribute values was determined by ranking the corresponding Shapley values at the utterance level, followed by averaging these rankings across the evaluation set and classes. The ASVspoof2019-det protocol was used for spoofing detection, while the ASVspoof2019-attr-17 protocol was employed for spoofing attack attribution.
\section{Experimental Results}\label{sec:exp-results}

In this section, we present the results, beginning with the baseline systems on two downstream tasks (spoofing detection and attack attribution) with CM embeddings, followed by the performance of the probabilistic attribute extractors and the downstream tasks using the proposed probabilistic attribute embeddings. Lastly, we analyze how the attributes contribute to decision-making using Shapley values with the best-performed system.

\subsection{Performance of original (raw) CM embeddings}

\textbf{Spoofing detection with CM embeddings.} Since our proposed framework uses spoofing countermeasure embeddings as the input features for later processing steps, we begin by evaluating the performance of these embedding extractors. To this end, we use the ASVspoof2019 standard spoofing detection protocol (ASVspoof2019-det) 
to evaluate the AASIST, RawBoost-AASIST, and SSL-AASIST models. We first report the spoofing detection performance of the original standalone `end-to-end' models. Using AASIST, the reproduced result aligns with the reported performance in~\cite{jung2022aasist}, achieving an EER of $0.83\%$. In the literature, RawBoost-AASIST and SSL-AASIST were evaluated on the ASVspoof2021 dataset in~\cite{tak2022rawboost} and~\cite{tak2022automatic}, respectively. Since the training partitions of ASVspoof2019 and ASVspoof2021 are identical, we evaluated these models on the ASVspoof2019 evaluation set for comparison. The obtained EERs were $1.59\%$ for RawBoost-AASIST and $0.22\%$ for SSL-AASIST. This performance trend remains consistent when the models are used as embedding extractors with auxiliary classifiers (DT, LR, and SVM), as summarized in Table~\ref{SD_asvspoof}. Regardless of the back-end classifier, SSL-AASIST embeddings consistently outperform RawBoost-AASIST and AASIST embeddings. Furthermore, for all embeddings, the LR classifier demonstrates superior spoofing detection performance compared to SVM and DT classifiers, achieving the highest balanced accuracy of $99.90\%$ and the lowest EER of $0.22\%$. Notably, the EER of the best-performing standalone CM system ($0.22\%$) matches the performance when SSL-AASIST is used as an embedding extractor followed by the LR classifier. This indicates that there is no significant performance loss whether the CM model is used directly or as an embedding extractor with the optimal auxiliary classifier.

\begin{table}[]
\centering
\caption{Spoofing detection performance of the CM embeddings in eval-set of  ASVspoof2019-det protocol. DT, LR, and SVM refer to decision tree, logistic regressing, and support vector machine respectively. The boldface signifies the best performance.}
\label{SD_asvspoof}
\begin{tabular}{|l|c|c|cccccc|}
\hline
\multirow{3}{*}{Protocol}                               & \multirow{3}{*}{\begin{tabular}[c]{@{}c@{}}CM Embedding\\ Extractor\end{tabular}} & Original CM   & \multicolumn{6}{c|}{CM as an embedding extractor}                                                                                                                     \\ \cline{3-9} 
                                                        &                                                                                   & EER           & \multicolumn{3}{c|}{Balanced accuracy}                                                        & \multicolumn{3}{c|}{EER}                                              \\ \cline{3-9} 
                                                        &            
                                                        & -             & \multicolumn{1}{c|}{DT}    & \multicolumn{1}{c|}{LR}    & \multicolumn{1}{c|}{SVM}            & \multicolumn{1}{c|}{DT}   & \multicolumn{1}{c|}{LR}   & SVM           \\ \hline
\multicolumn{1}{|c|}{\multirow{3}{*}{ASVspoof2019-det}} & AASIST~\cite{jung2022aasist}                                                                            & 0.83          & \multicolumn{1}{c|}{98.31} & \multicolumn{1}{c|}{98.69} & \multicolumn{1}{c|}{98.53}          & \multicolumn{1}{c|}{3.12} & \multicolumn{1}{c|}{0.92} & 1.13          \\ \cline{2-9} 
\multicolumn{1}{|c|}{}                                  & RawBoost-AASIST                                                                   & 1.59          & \multicolumn{1}{c|}{97.80} & \multicolumn{1}{c|}{98.27} & \multicolumn{1}{c|}{98.19}          & \multicolumn{1}{c|}{2.94} & \multicolumn{1}{c|}{1.47} & 1.55          \\ \cline{2-9} 
\multicolumn{1}{|c|}{}                                  & SSL-AASIST                                                                        & \textbf{0.22} & \multicolumn{1}{c|}{99.74} & \multicolumn{1}{c|}{\textbf{99.90}} & \multicolumn{1}{c|}{99.78} & \multicolumn{1}{c|}{0.34} & \multicolumn{1}{c|}{0.22} & \textbf{0.22} \\ \hline
\end{tabular}
\end{table}

\textbf{Attack attribution with CM embeddings.} For attack attribution, ASVspoof2019-attr-2 protocol is used for all three types of embeddings (AASIST, RawBoost-AASIST, and SSL-AASIST) and three back-end classifiers. 
The 
results are summarized in Table~\ref{tab:SAA_ecm}. 
Regardless of the back-end classifier, the performance of AASIST embeddings surpasses that of RawBoost and SSL-AASIST embeddings. Furthermore, across all embeddings, the highest balanced accuracy was achieved using the SVM classifier, while the lowest EER was obtained with the LR classifier.  The highest balanced accuracy and lowest EER achieved are $99.09\%$ and $0.34\%$, respectively.

\begin{figure}[h]
    \centering
    {\includegraphics[height= 150pt,width=320pt]{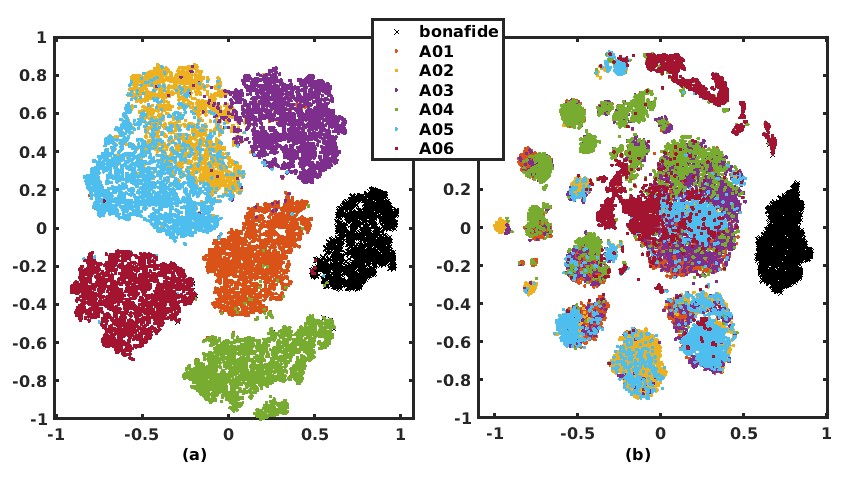}}
    \caption{The t-SNE projection of spoof CM embeddings obtained from (a) AASIST, and (b) SSL-AASIST systems, respectively.}
    \label{tsne_1}
  
\end{figure}

While SSL-AASIST shows better performance in spoofing detection, its attribution performance, with a balanced accuracy of $94.09\%$, is lower compared to AASIST. 
To investigate this, we 
display attack-specific \emph{t-distributed stochastic neighbor embedding} (t-SNE)~\cite{van2008visualizing} projections of 
AASIST and SSL-AASIST embedding in Figure~\ref{tsne_1} 
on the 
development data. 
We make two observations. First, the bonafide samples 
form distinct clusters in both embedding spaces, demonstrating clear separation from most attacks. Second, the attacks are more overlapped with each other in the SSL-AASIST embedding space. 
This is likely due to the SSL model being pre-trained on a different, out-of-domain dataset and then fine-tuned on ASVspoof2019 for spoofing detection. While this setup enhances generalization in spoofing detection across diverse attack types, it does dilute attack-specific characteristics. In contrast, AASIST is trained entirely on ASVspoof2019, which helps preserve detailed attack-specific information in its embeddings.
Based on these findings, 
we evaluate 
attack attribution on the larger ASVspoof2019-attr-17 
protocol only with the vanilla AASIST embeddings (the last row of Table \ref{tab:SAA_ecm}). The highest balanced accuracy of $90.16\%$ and lowest EER of $2.11\%$ are obtained with SVM and logistic regression back-ends. As expected, the overall performance drops on this larger protocol. 

\begin{table}[]
\centering
\caption{Spoofing attack attribution performance of the CM embeddings in the eval-set of ASVspoof2019-attr-2 and ASVspoof2019-attr-17 protocol. DT, LR, and SVM refer to decision tree, logistic regressing, and support vector machine respectively. The boldface signifies the best performance. }
\label{tab:SAA_ecm}
\begin{tabular}{|c|c|ccc|ccc|}
\hline
\multirow{2}{*}{Protocol}            & \multirow{2}{*}{\begin{tabular}[c]{@{}c@{}}CM Embedding\\ Extractor\end{tabular}} & \multicolumn{3}{c|}{Balanced accuracy}                                   & \multicolumn{3}{c|}{EER}                                               \\ \cline{3-8} 
                                     &                                                                                   & \multicolumn{1}{c|}{DT}    & \multicolumn{1}{c|}{LR}    & SVM            & \multicolumn{1}{c|}{DT}    & \multicolumn{1}{c|}{LR}            & SVM  \\ \hline
\multirow{3}{*}{ASVspoof2019-attr-2} & AASIST                                                                            & \multicolumn{1}{c|}{94.68} & \multicolumn{1}{c|}{97.93} & \textbf{99.09} & \multicolumn{1}{c|}{1.77}  & \multicolumn{1}{c|}{\textbf{0.34}} & 0.39 \\ \cline{2-8} 
                                     & RawBoost-AASIST                                                                   & \multicolumn{1}{c|}{89.99} & \multicolumn{1}{c|}{97.04} & 97.35          & \multicolumn{1}{c|}{3.69}  & \multicolumn{1}{c|}{1.54}          & 1.14 \\ \cline{2-8} 
                                     & SSL-AASIST                                                                        & \multicolumn{1}{c|}{82.13} & \multicolumn{1}{c|}{92.70} & 94.09          & \multicolumn{1}{c|}{6.75}  & \multicolumn{1}{c|}{2.75}          & 5.40 \\ \hline
ASVspoof2019-attr-17                 & AASIST                                                                            & \multicolumn{1}{c|}{77.74} & \multicolumn{1}{c|}{89.27} & \textbf{90.16} & \multicolumn{1}{c|}{11.77} & \multicolumn{1}{c|}{\textbf{2.11}} & 4.18 \\ \hline
\end{tabular}
\end{table}

\subsection{Performance of Probabilistic Attribute Extractors}

\begin{figure}[h!]
    \centering
    \includegraphics[height= 200pt,width=430pt]{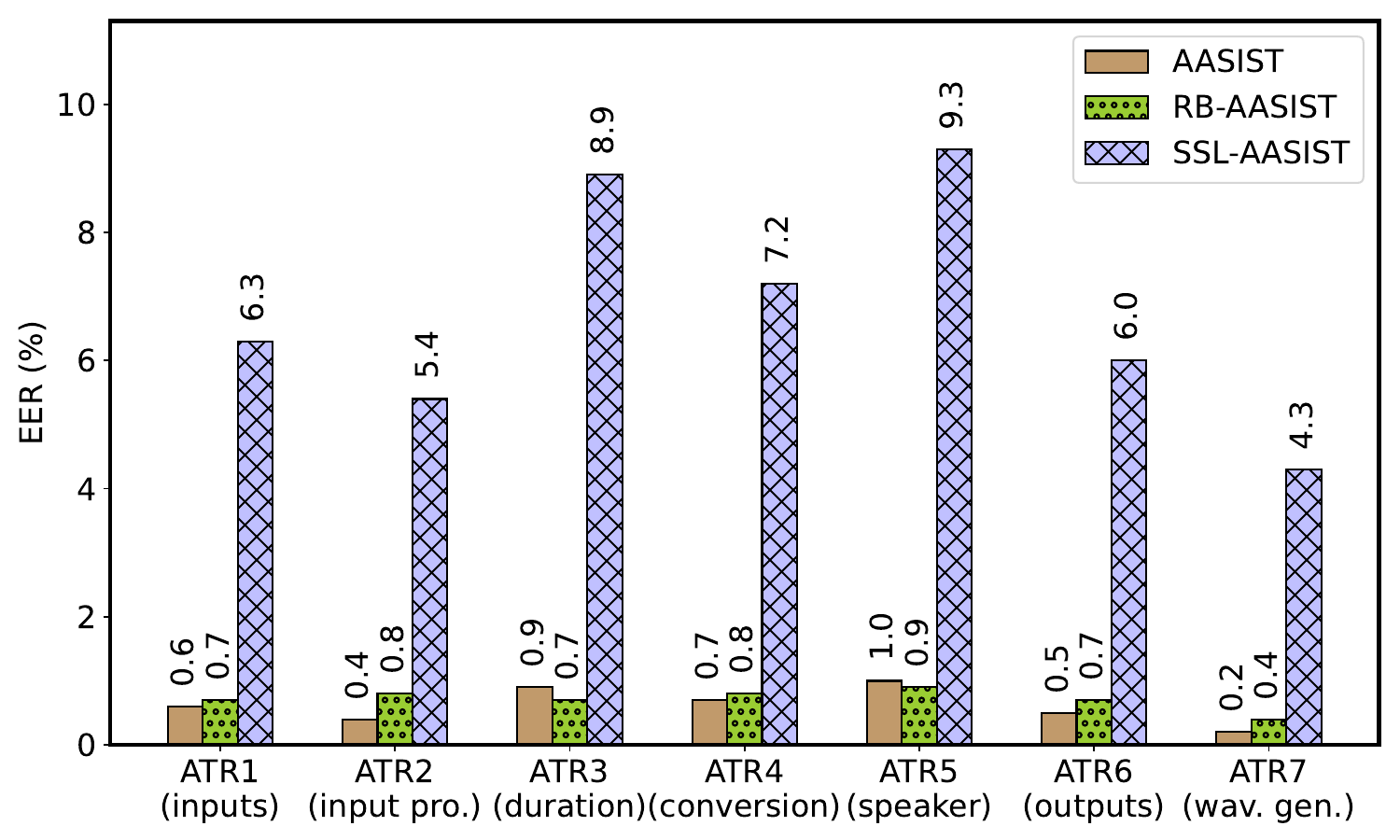}
    \caption{Probabilistic attribute extractor performance in development set using ASVspoof2019-det protocol for each attribute (ATR) shown in terms of EER, input pro.: input processor and wav. gen.: waveform generator. }
    \label{fig:avc_results}
\end{figure}

With the baseline embedding extractors now defined, we now turn our attention to the 
the  probabilistic attribute extractors.
Despite being discriminatively trained, the attribute extractors are viewed as feature extractors---i.e., part of our model selection strategy. Therefore, all the numbers reported in this subsection correspond to the \emph{development} parts of the included protocols.  
The results for the 
detection of attribute values 
are 
shown in Figure~\ref{fig:avc_results}. Concerning the choice of the CM embeddings, the findings are in line with the attribution performance in Table \ref{tab:SAA_ecm} 
---namely, SSL-AASIST being inferior (here, by an order of magnitude) to the two other types of embeddings.
As for the 
vanilla vs RawBoost variants of AASIST, 
the latter is generally inferior 
(except for the duration and speaker attributes). 

As for the relative performance for different attributes using 
vanilla AASIST, \texttt{waveform generation} achieves the lowest EER of $0.2\%$, followed by \texttt{input processor} ($0.4\%$), \texttt{outputs} ($0.5\%$), \texttt{inputs} ($0.6\%$), \texttt{conversion} ($0.7\%$), \texttt{duration} ($0.9\%$), and \texttt{speaker} ($1.0\%$). In \emph{absolute} terms all EERs are low ($\leq 1\%$), indicating the potential of a simple 'off-the-shelf' CM embeddings extractor followed up by MLP in extracting high-level spoofing attributes. Since this model structure 
is identical across all the attributes, 
the relative ordering reflects the extent to which traces of these attributes are present in the AASIST CM embeddings. In particular, waveform generator traces are most prominently present. 




On the basis of these findings, for 
the larger ASVspoof2019-attr-17 protocol, we only report the results with 
vanilla AASIST embeddings. 
The findings with this extended set of attribute values, presented in Table~\ref{tab:per-avc-asvspoof-ka}, 
are similar to 
the detection protocol results (Figure~\ref{fig:avc_results}):
the speaker and duration attributes performing somewhat worse compared to the others. Nonetheless, the \texttt{inputs} attribute now yields the lowest EER.
By comparing Figure~\ref{fig:avc_results} and Table~~\ref{tab:per-avc-asvspoof-ka}, the broad range\footnote{Technically, we should not compare the numbers one-to-one, given they are based on different evaluation protocols. Nonetheless, since the CM embeddings, classifier architectures and training recipes were kept as unified as possible, and with the same ASVspoof 2019 dataset being used for both experiments, broad comparison is deemed acceptable by the authors.} of EERs in the former is lower. 
This could be attributed to the increase in the number of attribute values, from $25$ (ASVspoof2019-det) to $50$ (ASVspoof2019-attr-17). Recall (Table \ref{tab:att-val-asvspoof}) that the number of attribute values (equivalently, the number of MLP outputs) is increased for all the seven attributes. We can \emph{a priori} expect this to decrease discrimination of values within each attribute.

\begin{figure}[h]
    \centering
    {\includegraphics[height= 230pt,width=420pt]{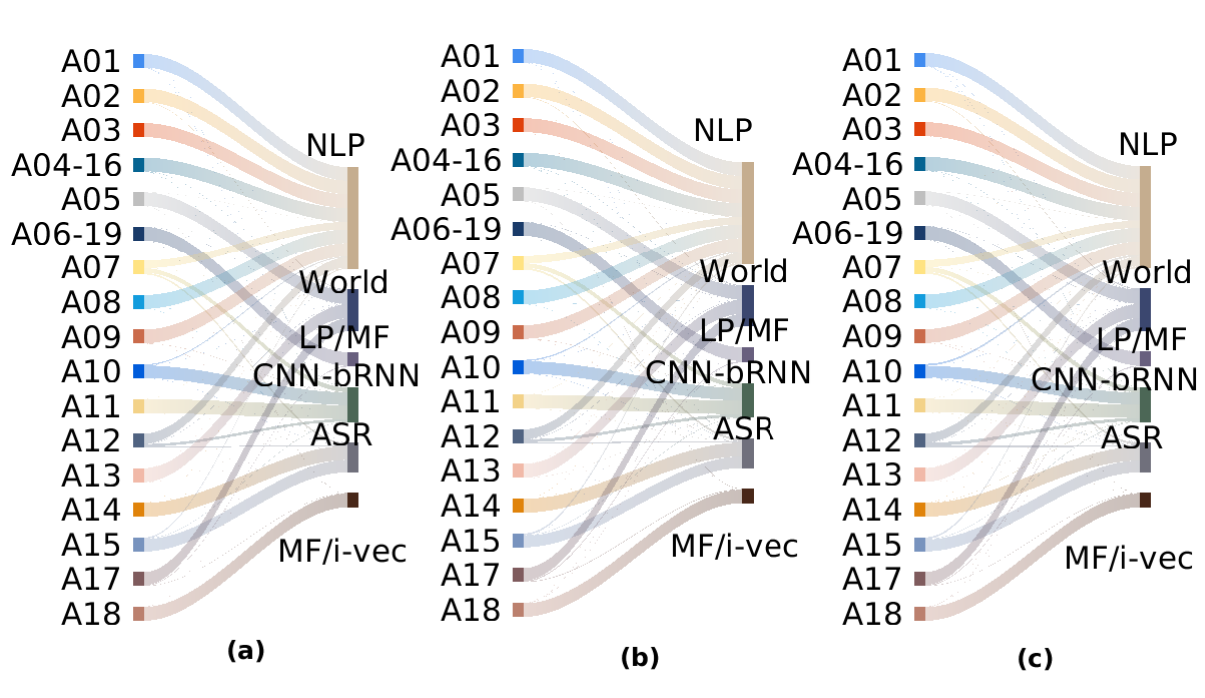}}
    \caption{The Sankey plot of attribute 2 (input processor) with AASIST embeddings in attribute value detection (a) training, (b) development, and (c) evaluation set of ASVspoof2019 attack attribution dataset, LP/MF: LPCC/MFCC, CNN-bRNN: CNN+bi-RNN and MF/i-vec: MFCC/ i-vector. The utterances specific to an attack are assigned to an attribute value based on the maximum prediction score from the probabilistic attribute extractor $\mathcal{F}^{\text{pae}}_{{\bm \rho}_{l}}$. This plot can be viewed as an alternative visual representation of the confusion matrix. The visualization provides an interpretation of how well the utterances belonging to a specific attack type can map to their corresponding attribute value using the trained probabilistic attribute extractor.  The corresponding EER for the input processor attribute is (a) $2.7\%$ on the training set, (b) $3.2\%$ on the development set, and (c)      $3.1\%$ on the evaluation set.}
    \label{fig:sankey_attr2}
  
\end{figure}

Our ground-truth attribute value vectors are categorical, with exactly one correct attribute value for a given spoofing attack and attribute. However, in practice, the learned mappings may exhibit inconsistencies or one-to-many relationships. Analyzing these confusions can provide insights into how well the probabilistic attribute extractor captures attribute-value associations, whether certain attribute values contribute more to misclassification, and whether these mappings are consistent across training, development, and evaluation partitions. For this analysis, we select the \texttt{input processor}\footnote{The \texttt{input processor} attribute is selected because it has a higher number of attribute values ($6$) compared to input ($3$). While other attributes may contain more than six values, including them would result in an excessively large and less accessible figure.} attribute, visualized using so-called \emph{Sankey plot}~\cite{schmidt2008sankey} in Figure~\ref{fig:sankey_attr2} for the training, development, and evaluation data. A Sankey plot represents visually the ``flow'' of data between different categories, illustrating how values transition from one state to another (here, the two states are attack labels and attribute values for \texttt{input processor} attribute). The width of each ``flow'' is proportional to the number of utterances mapped to a given attribute value. Since each attack is associated with exactly one correct attribute value, mappings \emph{without} splits indicate that the probabilistic attribute extractor has correctly learned the desired relationship between the attack label and its ground-truth attribute value. As observed, most utterances are correctly categorized for the \texttt{input processor} attribute, with only minor discrepancies in the utterances belonging to A07 and A12. Ideally, A07 should map to the \texttt{NLP} attribute value and A12 to \texttt{CNN+bi-RNN}; however, a small degree of confusion occurs for A07 between \texttt{CNN+bi-RNN} and \texttt{NLP}, and for A12 among \texttt{CNN+bi-RNN}, \texttt{NLP}, and \texttt{ASR} attribute values. Importantly, the Sankey plots remain largely consistent across the training, development, and evaluation sets, indicating that the probabilistic attribute extractor generalizes well, with the evaluation performance—showing stable attribute mappings similar to the training and development sets.



\begin{table}[]
\centering
\caption{Probabilistic attribute extractor performance in development set using ASVspoof2019-attr-17 protocol for each attribute(ATR) shown in terms of EER.}
\vspace{0.2 cm}
\label{tab:per-avc-asvspoof-ka}
\begin{tabular}{|c|c|c|c|c|c|c|c|}
\hline
 &
  \begin{tabular}[c]{@{}c@{}}ATR1\\ (inputs)\end{tabular} &
  \begin{tabular}[c]{@{}c@{}}ATR2\\ (input processor)\end{tabular} &
  \begin{tabular}[c]{@{}c@{}}ATR3\\ (duration)\end{tabular} &
  \begin{tabular}[c]{@{}c@{}}ATR4\\ (conversion)\end{tabular} &
  \begin{tabular}[c]{@{}c@{}}ATR5\\ (speaker)\end{tabular} &
  \begin{tabular}[c]{@{}c@{}}ATR6\\ (outputs)\end{tabular} &
  \begin{tabular}[c]{@{}c@{}}ATR7\\ (waveform gen)\end{tabular} \\ \hline
EER (\%) &
2.0   &
3.2  &
3.3   &
2.3   &
4.2   &
3.0   &
 2.9  \\ \hline
\end{tabular}
\end{table}

\subsection{Downstream Tasks using Probabilistic Attribute Embeddings}

\textbf{Spoofing detection with probabilistic attribute embeddings.} The performance of the two downstream tasks on the evaluation set(s), using the probabilistic attribute embeddings as input features are presented in Table~\ref{tab:comp_attr_cm}. For ease of reference, the table includes the performance achieved using both probabilistic attribute embeddings $\Vec{\rho}$ and raw CM embeddings $\Vec{e}_{\rm cm}$, replicated from Table~\ref{SD_asvspoof}. Regardless of the back-end classifiers, the performance of probabilistic attribute embeddings extracted from SSL-AASIST is superior to that of vanilla and RawBoost AASIST. This trend is consistent with the observations for CM embeddings in the spoofing detection task. Furthermore, with SSL-AASIST, the performance using probabilistic attribute embeddings $\Vec{\rho}$ with NB, DT, LR, and SVM classifiers in terms of balanced accuracy is $99.77\%$, $99.66\%$, $99.74\%$ and $99.74\%$, while in terms of EER, it is $0.22\%$, $0.54\%$, $0.23\%$ and $0.22\%$, respectively. These results indicate that performance, whether in terms of balanced accuracy or EER, remains relatively stable across auxiliary classifiers when using SSL-AASIST embeddings compared to the other methods. This suggests that the impact of auxiliary classifiers on the spoofing detection task is minimal if the extracted embeddings exhibit strong discrimination between spoofing and bonafide samples. Additionally, with SSL-AASIST, a slight performance drop is observed when using probabilistic attribute embeddings with the DT classifier. This suggests that DT classifiers are sensitive to probabilistic values. Compared to raw CM embeddings $\Vec{e}_{\rm cm}$, the performance in terms of balanced accuracy with DT, LR, and SVM classifiers is  $99.74\%$, $99.90\%$, and $99.78\%$, respectively, while in terms of EER, it is $0.34\%$, $0.22\%$, and $0.22\%$. These findings indicate that, regardless of the auxiliary classifier used, spoofing detection performance remains nearly identical when using either CM embeddings or explainable probabilistic attribute embeddings.

\begin{table}[h]
\centering
\caption{Performance comparison of spoofing detection: probabilistic attribute embeddings vs. CM embeddings on eval-set of ASVspoof2019-det protocol. DT, LR, and SVM refer to decision tree, logistic regressing, and support vector machine respectively. The boldface signifies the best performance.}
\label{tab:comp_attr_cm}
\resizebox{\columnwidth}{!}{%
\begin{tabular}{|c|c|ccccccc|ccccccc|}
\hline
\multirow{3}{*}{Protocol} &
  \multirow{3}{*}{\begin{tabular}[c]{@{}c@{}}CM Embedding\\ Extractor\end{tabular}} &
  \multicolumn{7}{c|}{Balanced accuracy} &
  \multicolumn{7}{c|}{EER} \\ \cline{3-16} 
 &
   &
  \multicolumn{1}{c|}{NB} &
  \multicolumn{2}{c|}{DT} &
  \multicolumn{2}{c|}{LR} &
  \multicolumn{2}{c|}{SVM} &
  \multicolumn{1}{c|}{NB} &
  \multicolumn{2}{c|}{DT} &
  \multicolumn{2}{c|}{LR} &
  \multicolumn{2}{c|}{SVM} \\ \cline{3-16} 
 &
   &
  \multicolumn{1}{c|}{$\Vec{\rho}$} &
  \multicolumn{1}{c|}{$\Vec{e}_{\text{cm}}$} &
  \multicolumn{1}{c|}{$\Vec{\rho}$} &
  \multicolumn{1}{c|}{$\Vec{e}_{\text{cm}}$} &
  \multicolumn{1}{c|}{$\Vec{\rho}$} &
  \multicolumn{1}{c|}{$\Vec{e}_{\text{cm}}$} &
  $\Vec{\rho}$ &
  \multicolumn{1}{c|}{$\Vec{\rho}$} &
  \multicolumn{1}{c|}{$\Vec{e}_{\text{cm}}$} &
  \multicolumn{1}{c|}{$\Vec{\rho}$} &
  \multicolumn{1}{c|}{$\Vec{e}_{\text{cm}}$} &
  \multicolumn{1}{c|}{$\Vec{\rho}$} &
  \multicolumn{1}{c|}{$\Vec{e}_{\text{cm}}$} &
  $\Vec{\rho}$ \\ \hline
\multirow{3}{*}{ASVspoof2019-det} &
  AASIST &
  \multicolumn{1}{c|}{84.48} &
  \multicolumn{1}{c|}{98.31} &
  \multicolumn{1}{c|}{95.83} &
  \multicolumn{1}{c|}{98.69} &
  \multicolumn{1}{c|}{57.88} &
  \multicolumn{1}{c|}{98.53} &
  54.94 &
  \multicolumn{1}{c|}{21.19} &
  \multicolumn{1}{c|}{3.12} &
  \multicolumn{1}{c|}{4.47} &
  \multicolumn{1}{c|}{0.92} &
  \multicolumn{1}{c|}{6.10} &
  \multicolumn{1}{c|}{1.13} &
  10.21 \\ \cline{2-16} 
 &
  RawBoost-AASIST &
  \multicolumn{1}{c|}{81.50} &
  \multicolumn{1}{c|}{97.80} &
  \multicolumn{1}{c|}{97.04} &
  \multicolumn{1}{c|}{98.27} &
  \multicolumn{1}{c|}{55.83} &
  \multicolumn{1}{c|}{98.19} &
  55.83 &
  \multicolumn{1}{c|}{22.6} &
  \multicolumn{1}{c|}{2.94} &
  \multicolumn{1}{c|}{5.23} &
  \multicolumn{1}{c|}{1.47} &
  \multicolumn{1}{c|}{7.05} &
  \multicolumn{1}{c|}{1.55} &
  5.27 \\ \cline{2-16} 
 &
  SSL-AASIST &
  \multicolumn{1}{c|}{\textbf{99.77}} &
  \multicolumn{1}{c|}{99.74} &
  \multicolumn{1}{c|}{99.66} &
  \multicolumn{1}{c|}{\textbf{99.90}} &
  \multicolumn{1}{c|}{99.74} &
  \multicolumn{1}{c|}{99.78} &
  99.74 &
  \multicolumn{1}{c|}{\textbf{0.22}} &
  \multicolumn{1}{c|}{0.34} &
  \multicolumn{1}{c|}{0.54} &
  \multicolumn{1}{c|}{\textbf{0.22}} &
  \multicolumn{1}{c|}{0.23} &
  \multicolumn{1}{c|}{0.22} &
  0.22 \\ \hline
\end{tabular}%
}
\end{table}

\textbf{Spoofing attack attribution with probabilistic attribute embeddings.} The results for both attribution protocols are shown in Table~\ref{tab:comp_attr_cm_asvspoof_KA}.  For ease of comparison, the results with the raw CM embeddings $\Vec{e}_{\rm cm}$ (replicated from Table~\ref{tab:SAA_ecm}) are included. Regardless of the auxiliary classifiers, 
the probabilistic attribute embeddings extracted using AASIST outperform 
RawBoost-AASIST and SSL-AASIST embeddings, a result in line with the results with raw CM embeddings.
With the ASVspoof2019-attr-2 protocol, the performance on the evaluation set using probabilistic attribute embeddings extracted from AASIST embeddings with NB, DT, LR, and SVM in terms of balanced accuracy is $99.79\%$, $99.08\%$, $99.14\%$, and $99.14\%$, and in terms of EER is $0.28\%$, $0.91\%$, $0.42\%$, and $0.49\%$, respectively. This 
indicates that performance remains relatively stable across classifiers. Interestingly, the naive
Bayes classifier achieves the best performance. 

Using the larger ASVspoof2019-attr-17 
evaluation set, 
the performance of spoofing attack attribution using probabilistic attribute embeddings extracted from AASIST with NB, DT, LR, and SVM in terms of balanced accuracy is $89.67\%$, $39.27\%$, $90.23\%$, and $90.17\%$, and in terms of EER is $4.71\%$, $11.77\%$, $2.11\%$, and $3.07\%$, respectively. The decision tree (DT) classifier shows degraded performance. 
Additionally, the naive Bayes classifier achieves $89.67\%$ balanced accuracy and $4.71\%$ EER, providing competitive performance 
compared to the best-performing classifier (logistic regression). 

Overall, how do our proposed probabilistic embeddings compare with the raw CM embeddings? On ASVspoof2019-attr-2, regardless of the auxiliary classifiers, the 
probabilistic attribute embeddings perform comparably to, or in some cases better than, the raw CM embeddings. We observe similar trend also on the ASVspoof2019-attr-17 protocol 
(with the DT classifier being an exception). 
These results confirm that the proposed probabilistic attribute embeddings perform as well as or better than raw CM embeddings, 
with the added benefits of 
explainability. It is also noteworthy that the dimensionality of the embeddings used for the downstream tasks is greatly reduced from the original ($160$) to $25$ (in the `det' and `attr-2' protocols) and $50$ (in the `attr-17' protocol). In this sense, our proposed probabilistic approach achieves better compaction of information between the audio waveforms and the class labels.

\begin{table}[h]
\centering
\caption{Performance comparison of spoofing attack attribution on eval-set of ASVspoof2019-attr-2 and ASVspoof2019-attr-17 protocols: probabilistic attribute embeddings vs. CM embeddings. DT, LR, and SVM refer to decision tree, logistic regressing, and support vector machine respectively. The boldface signifies the best performance.}
\label{tab:comp_attr_cm_asvspoof_KA}
\resizebox{\columnwidth}{!}{%
\begin{tabular}{|c|c|ccccccc|ccccccc|}
\hline
\multirow{3}{*}{Protocol} &
  \multirow{3}{*}{\begin{tabular}[c]{@{}c@{}}CM Embedding\\ Extractor\end{tabular}} &
  \multicolumn{7}{c|}{Balanced accuracy} &
  \multicolumn{7}{c|}{EER} \\ \cline{3-16} 
 &
   &
  \multicolumn{1}{c|}{NB} &
  \multicolumn{2}{c|}{DT} &
  \multicolumn{2}{c|}{LR} &
  \multicolumn{2}{c|}{SVM} &
  \multicolumn{1}{c|}{NB} &
  \multicolumn{2}{c|}{DT} &
  \multicolumn{2}{c|}{LR} &
  \multicolumn{2}{c|}{SVM} \\ \cline{3-16} 
 &
   &
  \multicolumn{1}{c|}{$\Vec{\rho}$} &
  \multicolumn{1}{c|}{$\Vec{e}_{\text{cm}}$} &
  \multicolumn{1}{c|}{$\Vec{\rho}$} &
  \multicolumn{1}{c|}{$\Vec{e}_{\text{cm}}$} &
  \multicolumn{1}{c|}{$\Vec{\rho}$} &
  \multicolumn{1}{c|}{$\Vec{e}_{\text{cm}}$} &
  $\Vec{\rho}$ &
  \multicolumn{1}{c|}{$\Vec{\rho}$} &
  \multicolumn{1}{c|}{$\Vec{e}_{\text{cm}}$} &
  \multicolumn{1}{c|}{$\Vec{\rho}$} &
  \multicolumn{1}{c|}{$\Vec{e}_{\text{cm}}$} &
  \multicolumn{1}{c|}{$\Vec{\rho}$} &
  \multicolumn{1}{c|}{$\Vec{e}_{\text{cm}}$} &
  $\Vec{\rho}$ \\ \hline
\multirow{3}{*}{ASVspoof2019-attr-2} &
  AASIST &
  \multicolumn{1}{c|}{\textbf{99.79}} &
  \multicolumn{1}{c|}{94.68} &
  \multicolumn{1}{c|}{99.08} &
  \multicolumn{1}{c|}{97.93} &
  \multicolumn{1}{c|}{99.14} &
  \multicolumn{1}{c|}{\textbf{99.09}} &
  99.14 &
  \multicolumn{1}{c|}{\textbf{0.28}} &
  \multicolumn{1}{c|}{1.77} &
  \multicolumn{1}{c|}{0.91} &
  \multicolumn{1}{c|}{\textbf{0.34}} &
  \multicolumn{1}{c|}{0.42} &
  \multicolumn{1}{c|}{0.39} &
  0.49 \\ \cline{2-16} 
 &
  RawBoost-AASIST &
  \multicolumn{1}{c|}{99.15} &
  \multicolumn{1}{c|}{89.99} &
  \multicolumn{1}{c|}{96.43} &
  \multicolumn{1}{c|}{97.04} &
  \multicolumn{1}{c|}{97.41} &
  \multicolumn{1}{c|}{97.35} &
  97.20 &
  \multicolumn{1}{c|}{0.82} &
  \multicolumn{1}{c|}{3.69} &
  \multicolumn{1}{c|}{1.46} &
  \multicolumn{1}{c|}{1.54} &
  \multicolumn{1}{c|}{1.22} &
  \multicolumn{1}{c|}{1.14} &
  0.97 \\ \cline{2-16} 
 &
  SSL-AASIST &
  \multicolumn{1}{c|}{98.75} &
  \multicolumn{1}{c|}{82.13} &
  \multicolumn{1}{c|}{91.64} &
  \multicolumn{1}{c|}{92.70} &
  \multicolumn{1}{c|}{94.50} &
  \multicolumn{1}{c|}{94.09} &
  94.35 &
  \multicolumn{1}{c|}{1.81} &
  \multicolumn{1}{c|}{6.75} &
  \multicolumn{1}{c|}{4.30} &
  \multicolumn{1}{c|}{2.75} &
  \multicolumn{1}{c|}{1.96} &
  \multicolumn{1}{c|}{5.40} &
  2.00 \\ \hline
ASVspoof2019-attr-17 &
  AASIST &
  \multicolumn{1}{c|}{89.67} &
  \multicolumn{1}{c|}{77.74} &
  \multicolumn{1}{c|}{39.27} &
  \multicolumn{1}{c|}{89.27} &
  \multicolumn{1}{c|}{\textbf{90.23}} &
  \multicolumn{1}{c|}{\textbf{90.16}} &
  90.17 &
  \multicolumn{1}{c|}{4.71} &
  \multicolumn{1}{c|}{11.77} &
  \multicolumn{1}{c|}{19.31} &
  \multicolumn{1}{c|}{\textbf{2.11}} &
  \multicolumn{1}{c|}{\textbf{2.07}} &
  \multicolumn{1}{c|}{4.18} &
  3.07 \\ \hline
\end{tabular}%
}
\end{table}

\subsection{Explanability of Probabilistic attribute embeddings}
\begin{figure}[h!]
    \centering
{\includegraphics[width=\linewidth]{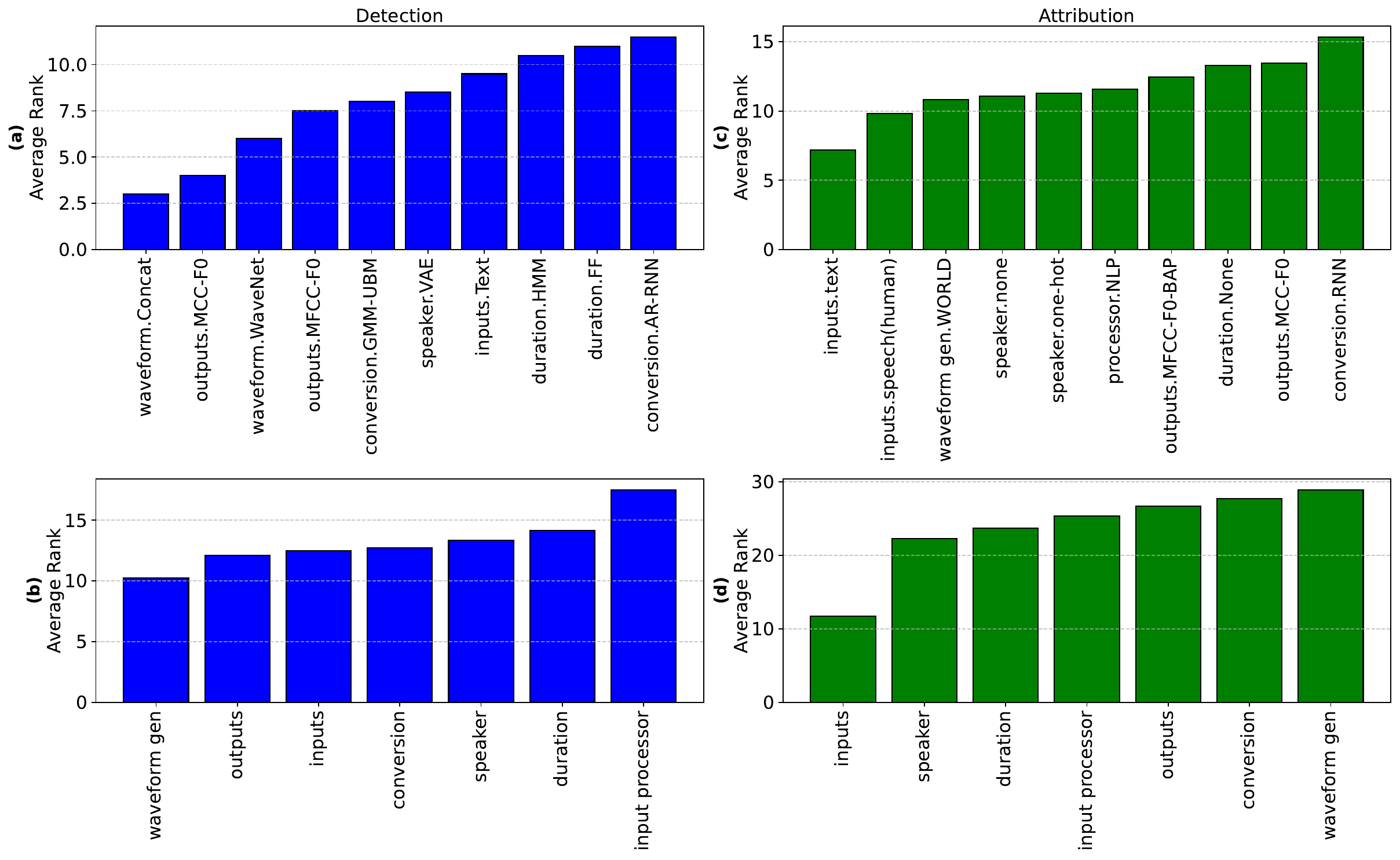}}
    \caption{Average rankings of Shapley value across utterances and classes for: (a) spoofing detection based on the top $10$ attribute values, (b) spoofing detection based on attributes, (c) spoofing attack attribution based on the top $10$ attribute values, and (d) spoofing attack attribution based on attributes. A lower rank indicates a higher contribution to decision-making.}
    \label{fig:shap_ranked}
  
\end{figure}

As our final analysis, we assess the relative importance of attribute values in probabilistic attribute embeddings to decision-making in downstream tasks using Shapley values~\cite{scott2017unified}. To analyze the relative importance of each attribute value, we selected the best-performing classifier for each downstream task: naive Bayes for the detection and logistic regression for the attribution task.  The contributions in terms of ranking of these attribute values and attributes on the eval-sets are illustrated in Figure~\ref{fig:shap_ranked} (a, b) for detection and  (c, d) for attribution. The overall ranking of each attribute is computed as the average of its individual attribute value rankings.

 Figure~\ref{fig:shap_ranked}~(a) highlights that \texttt{waveform gen.concat} (where "waveform gen" is the attribute and "concat" is a specific value), \texttt{outputs.MCC-F0}, and \texttt{waveform gen.WaveNet} are the top three attribute values that predominantly contribute to predictions in the spoofing detection task. When aggregated at the attribute level, \texttt{waveform gen}, \texttt{outputs}, and \texttt{inputs} emerge as the most influential attributes in spoofing detection. Similarly, Figure~\ref{fig:shap_ranked}~(c) shows that \texttt{input.text}, \texttt{input.speech (human)}, \texttt{waveform gen.WORLD}, and \texttt{speaker.one-hot} are among the top attribute values influencing predictions in the spoofing attack attribution task. In terms of aggregated contribution, the attributes \texttt{inputs}, \texttt{speaker}, and \texttt{duration} are the most dominant in attribution.

We observe that, in spoofing detection, \texttt{waveform gen}, \texttt{outputs}, and \texttt{inputs} attributes emerge as key factors in decision-making, suggesting that variations in these attribute values exhibit distinct distributions between spoofed and bonafide utterances, making them highly discriminative. This further indicates that the synthesis process might introduce artifacts linked to these attributes' design choices, which help differentiate spoofed speech from bonafide speech. Similarly, for spoofing attack attribution,  \texttt{inputs},  \texttt{speaker}, and  \texttt{duration} attributes play a crucial role in distinguishing between different spoofing attacks. This suggests that the specific design choices of these attributes might introduce distinct artifacts in the synthesized speech signals, providing valuable cues for effective spoofing attack attribution. These insights can be leveraged in the future to deepen our understanding of the spoofed speech generation process, enabling more transparent and informed decision-making. For further insight, class-wise rankings of Shapley values for all attributes and their values are available on our public GitHub\footnote{\url{https://github.com/Manasi2001/Spoofed-Speech-Attribution/tree/main/SHAP-plots}} repository. We warmly encourage interested readers to explore these resources for a more detailed understanding.

\section{A Broader Outlook --- Towards Generalizable Spoofed Speech Attributes}

Before concluding, the authors wish to pause for a moment to link our work to a broader context. Whereas our focus has been on discrete probabilistic characterization of spoofed speech, 
the initial inspirations (beyond the studies already cited in Section \ref{Sec:related-work}) originate from abstract descriptions of bonafide (real human) speech. 
Speech researchers are familiar to speech characterization based on abstract, discrete presentations. Spoken language consists of words; which in turn consist of syllables; which in turn consist of elementary speech sounds. These abstract elementary speech sounds can be described at phonemic (language-specific) and phonetic (language-independent) levels, and they form the basis for standardized speech transcription systems, such as the \emph{international phonetic alphabet} (IPA) \cite{ipa_chart}. In terms of the physical speech production process, speech sounds may further broken down into a more elementary set of binary-valued \emph{distinctive features} that relate to the manner and place of articulation; one may think the abstract phonetic representations to have a unique binary-valued manner and place characterization vector. While these abstract descriptions remain pivotal in linguistic descriptions of spoken language, they also find use in speech technology. For instance, the authors in \cite{Lee2013-information-extraction} proposed to divide-and-conquer the complex task of \emph{automatic speech recognition} (ASR) into hierarhically-organized, simpler sub-tasks. Their approach leveraged a bank of probabilistic speech attribute detectors, each optimized to estimate the degree to which a specific elementary speech property (e.g. nasality or silence) is present. Using evidence combination, the posterior probabilities of lower-level attributes are combined to form higher-level descriptions, such as \emph{phonetic posteriorgrams} \cite{Fousek2006-hierarchical} or word lattices. As opposed to spectrogram reading that requires substantial expertise and practice, intepreting a phonetic posteriorgram---display of phone posteriors as a function of time---could be argued to be more explainable. 

While the above approaches use discrete probabilistic speech characterization at the level of individual speech segments, 
important for tasks such as ASR \cite{Lee2013-information-extraction} and VC \cite{Sun2016-phonetic-posteriorgram-VC}, we may also characterize speech at the utterance (recording) level. 
Two widely studied utterance-level prediction task are speaker and language recognition. A key assumption 
in both is that the target class (speaker or language identity) remains fixed throughout a given utterance, justifying the use of a shared latent code across all the frames, typically referred to as a speaker (or language) embedding. A key research challenge has been to develop robust and compact 
speaker \cite{KinnunenLi2010,Hansen2015-speaker-rec-review,Bai2021-speaker-overview} or language \cite{Li2023-spoken-language-rec-overview} embeddings. 
A vast majority of these approaches are designed for fully-automated speaker or language recognition 
at scale, 
and without any humans in the loop. As a result, there has been little incentive to design speaker, language, or spoofed utterance representations that are interpretable. While some approaches (e.g. \cite{Lee2011-discrete,Bahari2014-GMM-weights}) 
use 
discrete probability embeddings to characterize speakers, they are obtained through unsupervised learning 
and yield a high-dimensional probability distribution, with no 
immediate meaning. One exception is  the binary attribute based speaker characterization approach \cite{Bonastre2011-binary-speaker,benamor24_interspeech,Imen2024-PhD-thesis} that inspired our work.

The two tasks considered in the present study, spoofing attack detection and attribution, have also been primarily addressed as utterance-level tasks (for exceptions where parts of utterances can be spoofed, see e.g. \cite{Zhang2023-partial-spoofing-IEEE-T,liu24m_interspeech}). This was also our purposefully selected focus, 
following the mainstream research in deepfake detection and biometric antispoofing. Nonetheless, a spoofed utterance is a time-domain signal where different parts of the utterance could contain different set of artifacts, some of them more informative of specific attributes. Therefore, one clear future direction is to study probabilistic 
attribution at the level of individual speech segments to obtain a more fine-grained representation, analogous 
to phonetic posteriorgrams \cite{Fousek2006-hierarchical} or manner/place extractors \cite{Lee2013-information-extraction}, but with relevant spoofed speech attributes used instead of phones or distinctive features. As argued in Section \ref{Sec:related-work}, however, 
such explainable characterization of artificial (spoofed) speech does not yet have similarly standardized, \emph{de facto} attributes. In the present work, we defined the attributes through specific metadata target labels available in a dataset, considered as a 'placeholder' in the absence of established theory for spoofed speech artifacts. The results presented in the previous section are promising and serve as an initial proof-of-concept. Readers interested in understanding how the probabilistic attributes behave when presented with bonafide speech samples are referred to the Appendix-2. As speech synthesis methodology evolves,  
it is foreseeable that explainable spoof attributes 
will likely be tied to specific datasets. The impact of unknown spoofing attacks on the proposed explainable framework is discussed in Appendix-3.  The longer-term aim, nonetheless, is to come up with 
generalizable attributes applicable to a wider array of spoofed speech generators. 


\section{Conclusion}

We proposed a novel framework for 
probabilistic 
characterization of spoofed speech, aimed at enhancing explainability and interpretability in 
spoofing detection and attack attribution (source tracing) tasks. Our initial findings 
are promising. In particular, they indicate that it is feasible to extract relatively low-dimensional probabilistic embeddings without compromising predictive performance relative to the use of original raw CM embeddings. 
For both 
tasks, the reported Shapley values 
revealed that \texttt{waveform generator}, \texttt{conversion model outputs}, and \texttt{inputs} attributes play an important 
in spoofing detection, whereas \texttt{inputs}, \texttt{speaker modeling}, and \texttt{duration}  are important in attribution. 
These encouraging results warrant further investigations both into the attribute extractors as well as the back-end classifiers. Additionally, we observed that SSL-AASIST embeddings perform better for spoofing detection, while AASIST embeddings are more effective for spoofing attack attribution, outperforming Raw-boost variant in their respective tasks, highlighting the complementary roles of self-supervised learning and data augmentation.

Our findings highlight the effectiveness of the CM embedding extractor and MLPs as a foundation for generating probabilistic attribute embeddings. While this serves as a strong starting point, we see opportunities for further improvements. Moving forward, we plan to refine the framework by optimizing the embeddings directly for classifying spoofing attacks and concerned attributes, to enhance their overall effectiveness. To ensure broader applicability across diverse scenarios, we aim to evaluate the framework on the recently available MLAAD \cite{muller2024mlaad} and ASVspoof5 corpora \cite{wang24_asvspoof}, fostering domain and language generalization. In the present study, focused on conceptual contribution, our intentional focus has been on the ASVspoof 2019 corpus which consists of high-quality audio materials. 
Generalization to other types of domains and attacks will be addressed in future work. 
Another important direction is to evaluate the robustness of the proposed approach under real-world conditions containing nuisance variations. These include, but are not limited to, 
speech codec or audio compression effects 
and additive noise. Future efforts will focus on systematically assessing and enhancing the framework’s resilience against these factors. Furthermore, a promising direction is the integration of spoofing detection and spoofing attack attribution within a unified framework. This would require careful design of evaluation protocols and metrics, and will be explored in future studies. Overall, our work represents a modest step toward developing explainable and interpretable frameworks for robust spoofing detection and spoofing attack attribution, with promising avenues for further enhancements.

\section*{Appendix-1: ML estimator for the probabilistic embedding features}\label{sec:appendix-ML-estimator}

Let us now detail how the score \eqref{eq:likelihood-score} can be computed in practice, and how the parameter estimates $\vec{\theta}_i$ can be obtained. Given that the probabilistic characterization vectors have typically low dimensionality relatively to the training sample size, we first approach the parameter estimation task through  \emph{maximum likelihood} (ML) estimation. First, to avoid notational clutter, we drop the raw waveform variable $\vec{x}$ since our back-end models have access to the probabilistic embeddings only, assumed to be 'sufficient statistics' for the attribution task. Further, since the  likelihood computation is identical for all the attacks, we drop the attack index $i$. The likelihood for one utterance, corresponding to the $\ell$th attribute set, is now written as $P(\vec{\rho}_\ell|\vec{\theta})$. 

To proceed, recall that we assume a discrete one-out-of-$M_\ell$ (or categorical) distribution for the attribute set $\ell$. To this end, let $\theta_{\ell,m} \mydef P(\rho_{\ell,m}=1)$ denote a scalar parameter---the probability for the $m$th element of $\vec{\rho}_\ell$ to be active (value 1). If we let $\vec{\theta}_\ell \mydef  (\theta_{\ell,1},\dots\theta_{\ell,M_\ell})\transp$ to denote the parameters of the $\ell$th categorical distribution, the parameter vector of the full model (for a given attack) is $\vec{\theta}=\left(\vec{\theta}_1,\dots,\vec{\theta}_L\right)\transp=\left(\theta_{1,1},\theta_{1,2},\dots,\theta_{1,M_1},\theta_{2,1},\theta_{2,2},\dots,\theta_{\ell,m},\dots,\theta_{L,M_L}\right)\transp$. Each of the parameter vectors $\vec{\theta}_\ell$ represents a discrete probability mass function corresponding to the $\ell$th attribute. This model has one parameter per attribute value, with a total of $\sum_{\ell=1}^L M_\ell$ parameters.

The likelihood for one observation $\vec{\rho}_\ell$ can now be written as~\cite{bishop2006pattern},
    \begin{equation}
        P(\vec{\rho}_\ell|\vec{\theta}) = \prod_{m=1}^{M_\ell} \left({\theta_{\ell,m}}\right)^{\rho_{\ell,m}}.
    \end{equation}
Now, given a training set $\mathcal{D}$ of $N$ independent observations denoted by $\mathcal{D}=\left\{\vec{\rho}_{\ell}^{(n)}\right\}_{n=1}^N$, 
the likelihood function for $\mathcal{D}$ becomes
    \begin{equation}
        P(\mathcal{D}|\vec{\theta})= \prod_{n=1}^N\prod_{\ell=1}^L \prod_{m=1}^{M_\ell} \left({\theta_{\ell,m}}\right)^{\rho_{\ell,m}^{(n)}} = \prod_{\ell=1}^L \prod_{m=1}^{M_\ell} \left(\prod_{n=1}^N \left({\theta_{\ell,m}}\right)^{\rho_{\ell,m}^{(n)}}\right)
        = \prod_{\ell=1}^L \prod_{m=1}^{M_\ell}  \left({\theta_{\ell,m}}\right)^{\sum_{n=1}^N\rho_{\ell,m}^{(n)}}
        = \prod_{\ell=1}^L \prod_{m=1}^{M_\ell} \theta_{\ell,m}^{S_{\ell,m}},
    \end{equation}
where $S_{\ell,m}\mydef \sum_{n=1}^N\rho_{\ell,m}^{(n)}$ denotes the number of training observations with the value 1 for attribute $m$ in attribute set $\ell$. The corresponding log-likelihood function is then
    \begin{equation}
        \log P(\mathcal{D}|\vec{\theta}) = \sum_{\ell=1}^L\sum_{m=1}^{M_\ell} S_{\ell,m}\log{\theta_{\ell,m}}.
    \end{equation}
Recall that the parameters for each attribute should represent probabilities of each of the individual attribute values being present. Thus, in order to find the maximum likelihood estimate $\hat{\vec{\theta}}^\text{ML}$ for the parameter vector $\vec{\theta}$, it is necessary to add probabilistic constraints of the form $\sum_{m=1}^{M_\ell} \theta_{\ell,m}=1$ for each $\ell=1,\dots,L$. By using the method of Lagrange multipliers, the objective function for ML parameter estimation now becomes
    \begin{equation}\label{eq:ML-estimate-for-categorical-distribution}
        \mathcal{L}(\vec{\theta},\vec{\lambda}) \mydef  \underbrace{\sum_{\ell=1}^L\sum_{m=1}^{M_\ell} S_{\ell,m}\log{\theta_{\ell,m}}}_{\text{log-likelihood}} + \underbrace{\sum_{\ell=1}^L \lambda_\ell\left(\sum_{m=1}^{M_\ell}\theta_{\ell,m} - 1 \right)}_{\text{the $L$ probabilistic constraints}},
    \end{equation}
where $\vec{\lambda}=\left(\lambda_1,\dots,\lambda_L\right)\transp$ denotes the vector of Lagrange multipliers, each responsible for constraining the corresponding subset of attribute values in attribute set $\ell$ to be normalized. Differentiating \eqref{eq:ML-estimate-for-categorical-distribution} with respect to the $\theta_{\ell,m}$ and setting the resulting expression to equal 0 yields $\hat{\theta}_{\ell,m}^\text{ML} = -S_{\ell,m}/{\lambda_\ell}$. The values for $\lambda_\ell, \ell=1,\dots,L$ can then be found by substituting this expression for each $\ell$ to the corresponding constraint equation $\sum_{m=1}^{M_\ell} \theta_{\ell,m}=1$, giving $\lambda_\ell = -\sum_{m=1}^{M_\ell} S_{\ell,m}$. The ML estimator therefore becomes 
    \begin{equation}
        \hat{\theta}_{\ell,m}^\text{ML} = \frac{S_{\ell,m}}{\sum_{m'=1}^{M_\ell} S_{\ell,m'}},\;\;\; \ell=1,\dots,L,\;m=1,\dots,M_\ell,
    \end{equation}
which is simply the relative frequency of training observations with the $m$th attribute within the $\ell$th attribute set being 1. As a data structure, the parameters for each of the spoofing attacks being modeled becomes a set of $L$ categorical distributions. The parameters of each attribute set can be estimated independently of the other attribute sets.
\section*{Appendix-2: Probabilistic attributes with bonafide utterances}

 Recall that all our probabilistic attribute detectors have been trained using deepfake speech only. Given that they are nonetheless used as features in a task (deepfake detection) that involves bonafide samples, one may ask \emph{what is the expected behavior of these features in response to bonafide speech?} 

 \begin{figure}[h!]
    \centering
    \includegraphics[height=300pt,width=460pt]{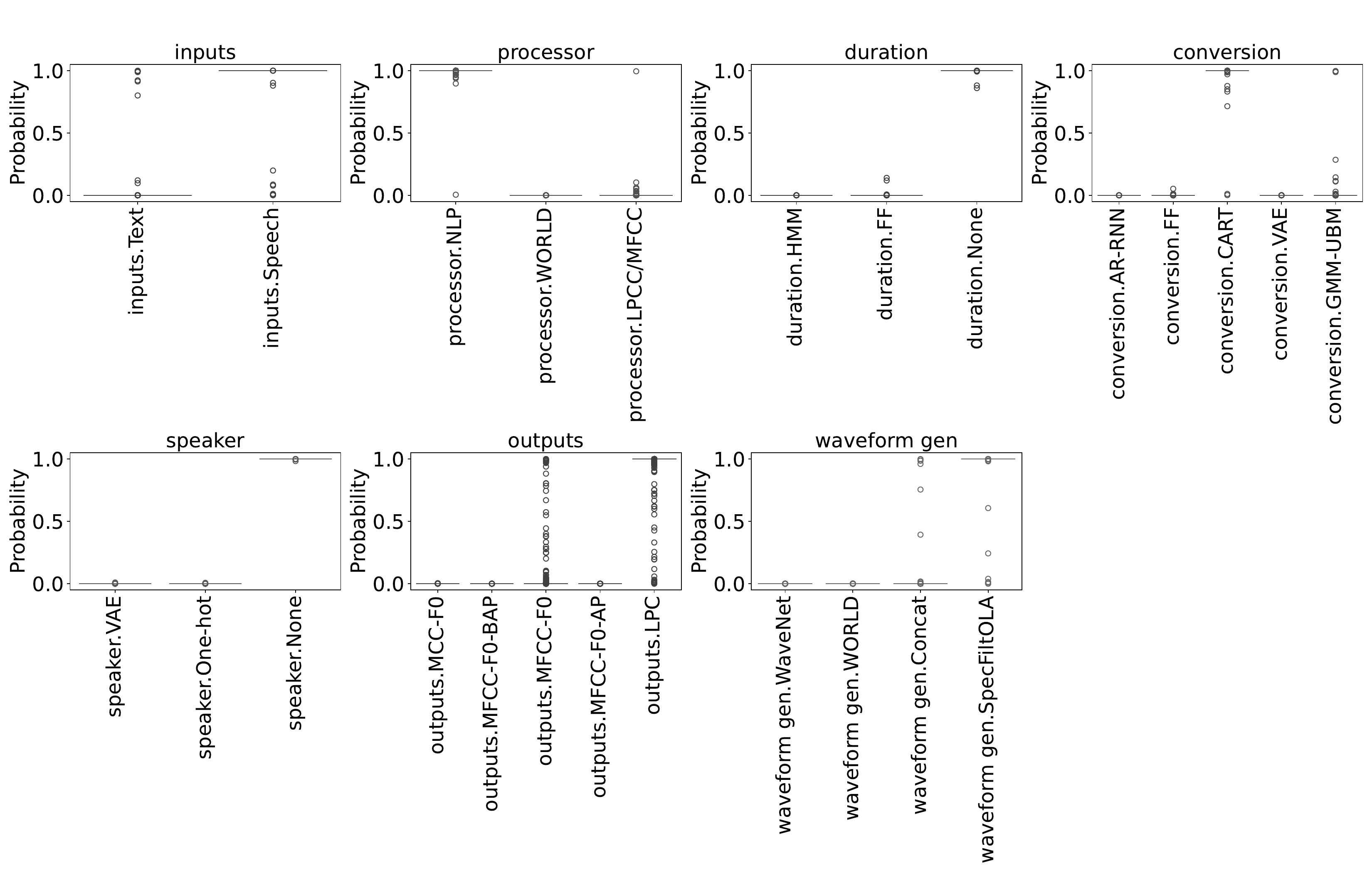}
    \caption{Distribution of probabilistic attribute values for the bonafide speech utterances.}
    \label{fig:bonafide}
\end{figure}

To explore this, we used the probabilistic attribute extractors trained with the ASVspoof2019-det protocol. Figure~\ref{fig:bonafide} presents the output distributions from each attribute detector when applied to bonafide samples from the evaluation set. Since the attribute–value pairs are inspired by characteristics of spoofed speech generation, our hypothesis is that bonafide utterances might be assigned to the attribute values that are closest to those seen in spoofed speech. In these cases, the prediction distributions are highly polarized, with probabilities concentrated near $0$ or $1$. For example, in the case of the \texttt{input} attribute, bonafide speech is predominantly assigned a value of $1$ for \texttt{Speech} and $0$ for \texttt{Text}. Similarly, for other attributes, bonafide speech is consistently assigned to specific values: \texttt{NLP} for \texttt{processor}, \texttt{None} for \texttt{duration} and \texttt{speaker}, \texttt{CART} for \texttt{conversion}, \texttt{LPC} for \texttt{output}, and \texttt{SpectFiltOLA} for \texttt{waveform}, with all other values receiving near-zero probabilities.

Interestingly, the \texttt{input} attribute shows that bonafide speech is more closely aligned with voice conversion (VC) attacks (where \texttt{input} is \texttt{speech}) than with text-to-speech (TTS) attacks (where \texttt{input} is \texttt{text}). Moreover, whenever \texttt{None} is a possible attribute value, bonafide speech tends to be assigned to it. These observations suggest that the attribute–value predictions for bonafide samples tend to cluster around specific values, potentially indicating a form of implicit alignment or resemblance between certain spoofed configurations and genuine speech.



In the future, we plan to expand the set of attributes to better explain bonafide speech by incorporating characteristics related to the speech production mechanism, and integrate these with the existing attributes to enhance interpretability of the decision-making process.
\section*{Appendix-3: Impact of Unknown Attacks on the Proposed Framework}
\begin{figure}[h!]
    \centering
    \includegraphics[height=430pt,width=480pt]{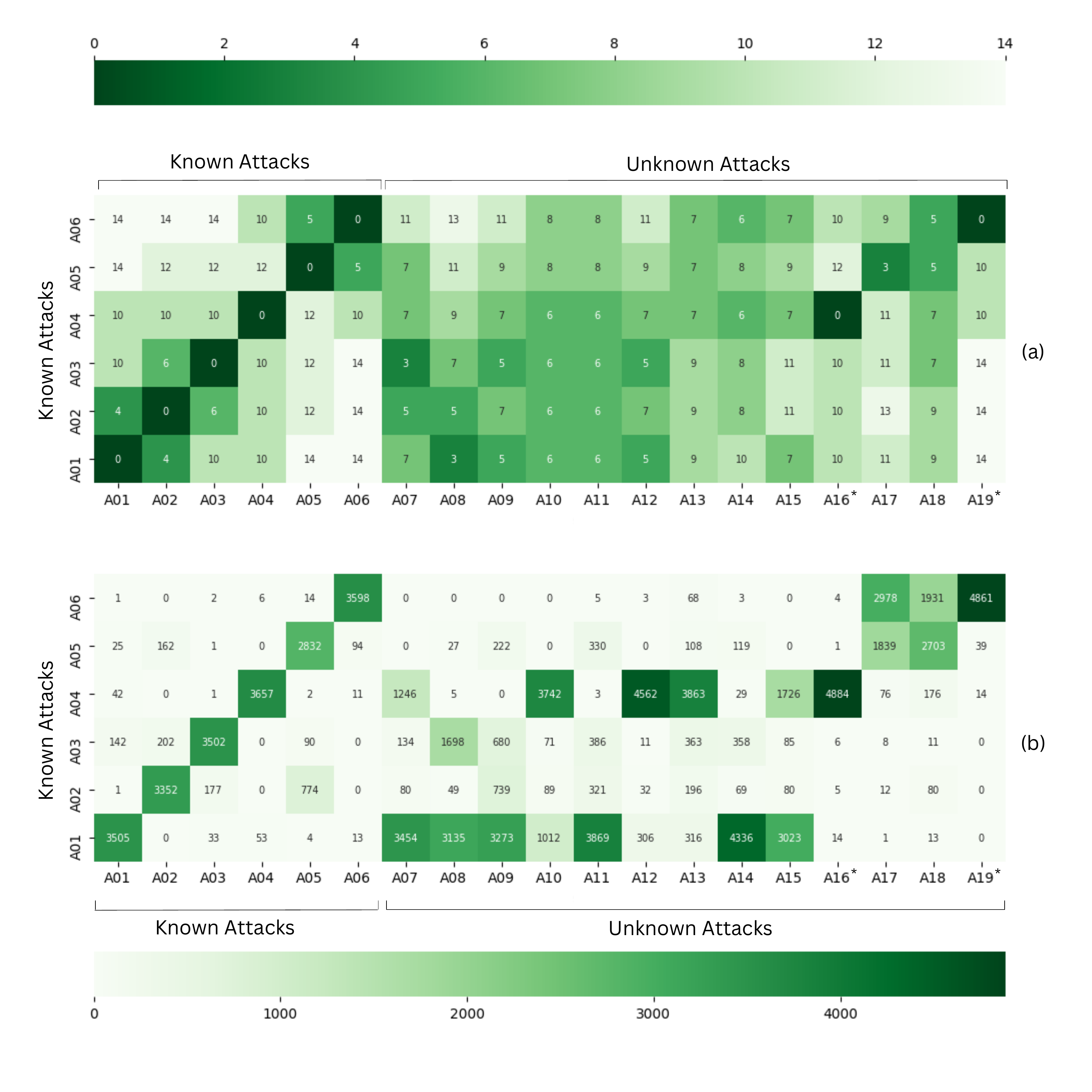}
    \caption{(a) Hamming distance and (b) confusion matrix. A16$^{*}$ and A19$^{*}$ are treated as known attacks, as their generation configurations are identical to A04 and A06, respectively. The color schemes for the Hamming distance and confusion matrix are intentionally inverted to enhance visualization, given the inverse relationship between the two. In the confusion matrix, A01–A06 correspond to spoofing attacks from the development set, while A07–A19 represent attacks from the evaluation set.}
    \label{fig:hamm_conf_plot}
\end{figure}

 The attribution experiments in this study have been focused on the closed-set set assumption, i.e. that all test time attacks are known at the training time. 
It is nonetheless relevant to also analyze the response of attribution methods to unseen attacks. To this end, we used the full evaluation set (excluding bonafide samples) from the original ASVspoof2019 protocol. 

Even if the unseen evaluation set attacks do not have a corresponding label in the training set, it is possible to quantify closeness (hence, confusability) of any pair of attacks through the distance of their ground-truth attribute vectors. For this purpose, we use 
%
Hamming distance of the 
one-hot encoded attribute vectors of known attacks (from the training and development sets) and those of unknown attacks (from the evaluation set). Hamming distance of 0 means that the pair of attacks are identical in terms of their ground-truth attributes. Using probabilistic attribute extractors trained on the training set of ASVspoof2019-attr-2, we classified evaluation set utterances—comprising unknown attacks—into the known attack classes. The resulting confusion matrix was analyzed to examine the relationship between unknown-to-known attack assignments and their corresponding Hamming distances. Our hypothesis is that unknown attacks are more likely to be assigned to known classes with smaller Hamming distances. The computed Hamming distances and the confusion matrix are presented in Figure~\ref{fig:hamm_conf_plot}.

It is evident that known attacks with a Hamming distance of zero exhibit clear 
correspondence, while the likelihood of correct attack assignment decreases as the Hamming distance increases. Extending this observation to unknown-to-known attack assignments, the trend generally holds, although not in all cases. This behavior suggests that improving the generalizability of both the embedding extractor and the probabilistic attribute extractor could enhance performance. In real-world scenarios—where novel spoofing attacks are continually emerging—this approach offers a principled way to assign new attacks to the most similar known classes, with the added benefit of explainability.

\section*{Acknowledgment}
The work has been partially supported by the Academy of Finland (Decision No. 349605, project "SPEECHFAKES"). The authors wish to acknowledge
CSC – IT Center for Science, Finland, for computational resources.

 \bibliographystyle{elsarticle-num} 
 \bibliography{cas-refs}

\end{document}